\documentclass[10pt,journal]{IEEEtran}
\usepackage[utf8]{inputenc} 
\usepackage[T1]{fontenc}    
\usepackage{hyperref}       
\usepackage{url}            
\usepackage{braket}
\usepackage{amsfonts}       
\usepackage{comment}
\usepackage{xcolor}

\newcommand{\ignore}[1]{}
\usepackage{fancyhdr}
\usepackage[normalem]{ulem}
\usepackage{microtype}
\usepackage{xspace}
\usepackage{amssymb}
\usepackage{pgfplots}
\pgfplotsset{compat=1.14}
\usetikzlibrary{patterns}
\usetikzlibrary{arrows}
\usepackage{xstring}
\usepackage[qm]{qcircuit}
\usepackage[boxed]{algorithm2e}
\usepackage{braket}
\usepackage{xcolor}
\usepackage{tabularx}
\usepackage{commath}
\usepackage{textcomp}
\usepackage[capitalise]{cleveref}
\usepackage{authblk}

\usepackage{physics}
\usepackage[]{subfig}

\usepackage{tikz}
\usetikzlibrary{automata, positioning, arrows, decorations.markings}
\tikzset{myptr/.style={decoration={markings,mark=at position 1 with %
    {\arrow[scale=2,>=stealth]{>}}},postaction={decorate}}} 
\newcommand{\colorQubit}{blue}
\newcommand{\colorQubitBB}{gray}
\newcommand{\colorQutrit}{orange!70!red}
\newcommand{\colorLightness}{60}
\newcommand{\colorQubitLight}{\colorQubit!\colorLightness}

\newcommand{\colorQutritLight}{\colorQutrit!\colorLightness}
\definecolor{brown}{RGB}{77, 51, 0}
\newcommand{\tSMTstar} {{T-SMT$^\bigstar$}}
\newcommand{\rSMTstar} {{R-SMT$^\bigstar$\xspace}}
\newcommand{\ibmnamefull} {{\em IBMQ16 Rueschlikon}}

\begin{document}
\title{Resource-Efficient Quantum Computing \\ by Breaking Abstractions}

\author[1]{Yunong Shi}
\author[1]{Pranav Gokhale}
\author[2]{Prakash Murali}
\author[1]{Jonathan M. Baker}
\author[1]{Casey Duckering}
\author[1]{Yongshan Ding}
\author[3]{Natalie C. Brown}
\author[5,6]{Christopher Chamberland}
\author[4]{Ali Javadi Abhari}
\author[4]{Andrew W. Cross}
\author[1]{David I. Schuster}
\author[1]{Kenneth R. Brown}
\author[2]{Margaret Martonosi}
\author[1]{Frederic T. Chong}

\affil[1]{University of Chicago}
\affil[2]{Princeton University}
\affil[3]{Duke University}
\affil[4]{IBM T.J. Watson Research Center}
\affil[5]{AWS Center for Quantum Computing}
\affil[6]{California Institute of Technology}

\maketitle

\begin{abstract}
Building a quantum computer that surpasses the computational power of its classical counterpart is a great engineering challenge. Quantum software optimizations can provide an accelerated pathway to the first generation of quantum computing applications that might save years of engineering effort. Current quantum software stacks follow a layered approach similar to the stack of classical computers, which was designed to manage the complexity. In this review, we point out that greater efficiency of quantum computing systems can be achieved by breaking the abstractions between these layers. We review several works along this line, including two hardware-aware compilation optimizations that break the quantum Instruction Set Architecture (ISA) abstraction and two error-correction/information-processing schemes that break the qubit abstraction. Last, we discuss several possible future directions.

\end{abstract}
\section{Introduction}

Quantum computing has recently transitioned from a theoretical prediction to a nascent technology. With development of Noisy Intermediate-Scale Quantum (NISQ) devices, cloud-based Quantum Information Processing (QIP) platforms with up to 53 qubits are currently accessible to the public. It also has been recently demonstrated by the {\it Quantum Supremacy} experiment on the Sycamore quantum processor, a 54-qubit quantum computing device manufactured by Google, that quantum computers can outperform current classical supercomputers in certain computational tasks \cite{supremacy}. These developments suggest that the  future of quantum computing is promising. Nevertheless, there is still a gap between the ability and reliability of current QIP technologies and the requirements of the first useful quantum computing applications.  The gap is mostly due to the presence of systematic errors including qubit decoherence,  gate errors, State Preparation And
Measurement (SPAM) errors. As an example, the best reported qubit decoherence time on a superconducting QIP platform is around 500$\mu$s (meaning that in 500$\mu$s, the probability of a logical 1 state staying unflipped drops to $1/e\approx 0.368$), the error rate of 2-qubit gates is around 1\%-5\% in a device, measurement error of a single qubit is between 2\%- 5\% \cite{ibmerror}. In addition to the errors in the elementary operations, emergent error modes such as crosstalk are reported to make significant contributions to the current noise level in quantum devices \cite{Sarovar2019DetectingCE,2018APSMARK39009L,alex2019characterizing}. With these sources of errors combined, we are only able to run quantum algorithms of very limited size on current quantum computing devices.
Thus, it will require tremendous efforts and investment to solve these engineering challenges and we cannot expect a definite timeline for its success.  Because of the uncertainties and difficulties in relying on hardware breakthroughs, it will also be crucial in the near term to close the gap  using higher-level quantum optimizations and software hardware co-design, which could maximally utilize noisy devices and potentially provide an accelerated pathway to real-world quantum computing applications.

\begin{figure}[t]
\centering
\includegraphics[width=0.32\textwidth]{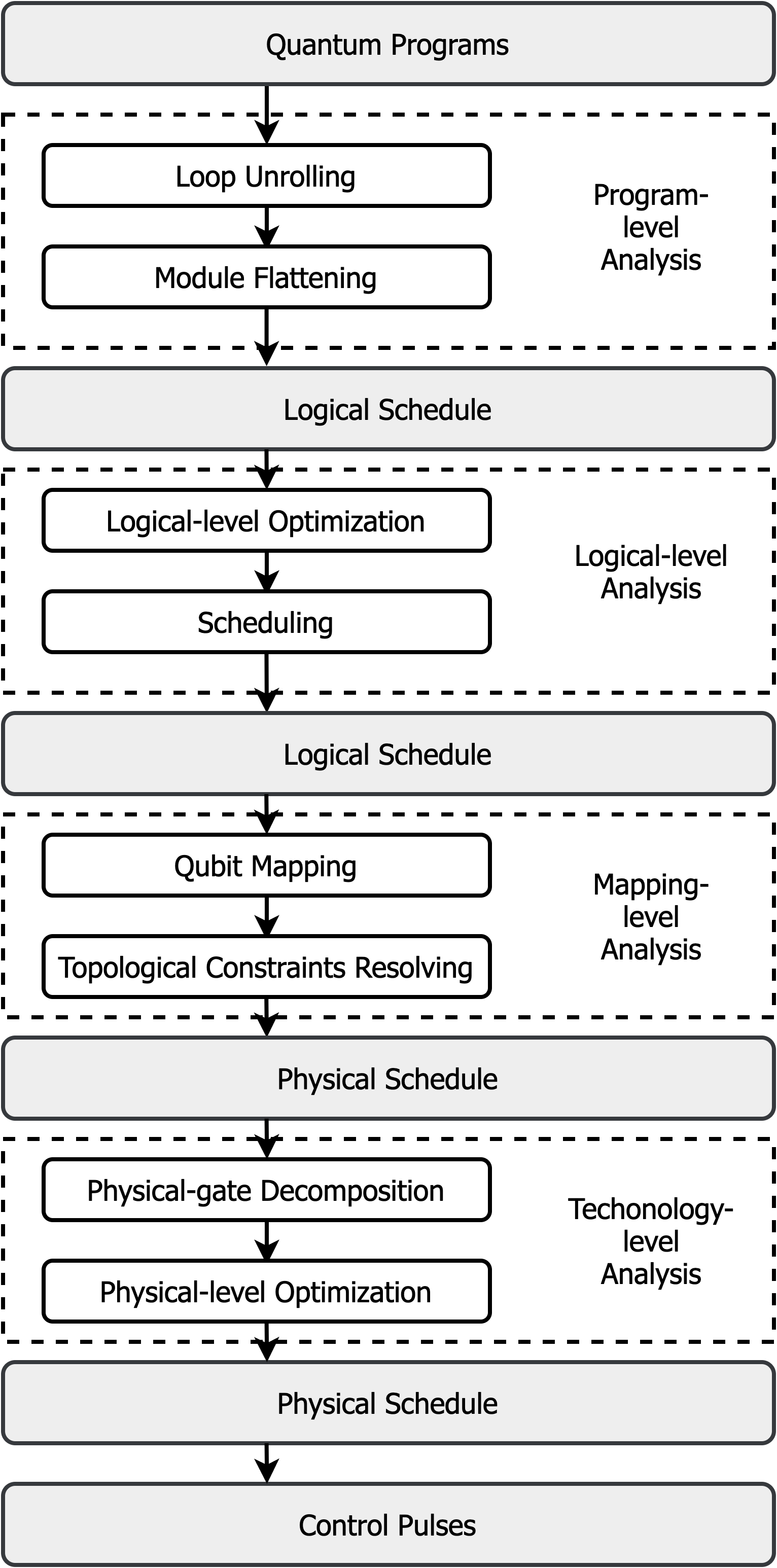}
\caption{The workflow of the quantum computing stack roughly followed by current programming environments ($e.g.,$ Qiskit, Cirq, ScaffCC) based on the quantum circuit model. }
\label{fig:layers}
\end{figure}

\begin{figure*}[t]
    \centering
    \includegraphics[width=0.98\textwidth]{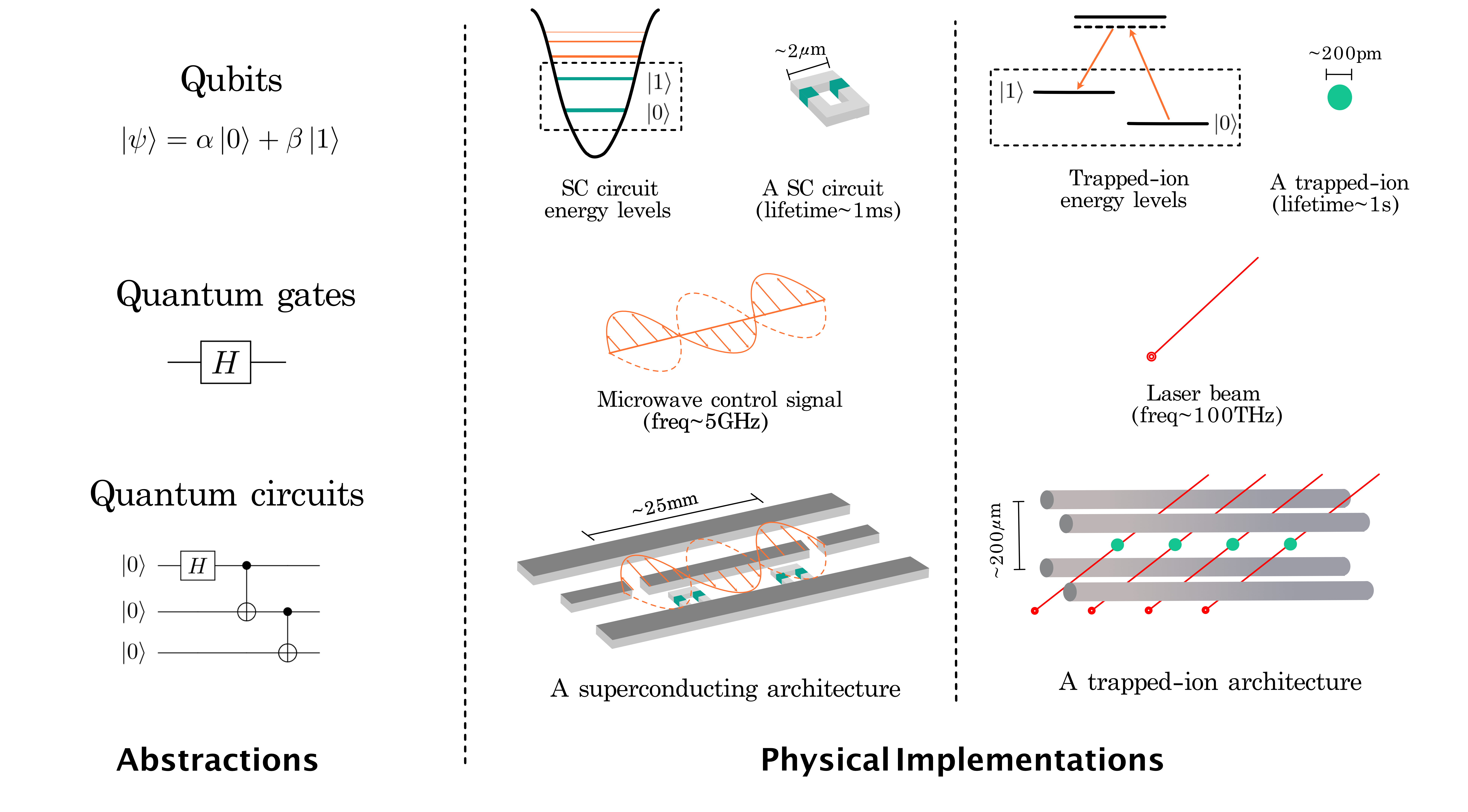}
    \caption{The same abstractions in the quantum computing stack on the logical level can be mapped to different physical implementations. Here we take the Superconducting (SC) QIP platform and the trapped ion QIP platform as examples of the physical implementations. (Left) In the quantum circuit model, both SC qubits and trapped-ion qubits are abstracted as 2-level quantum systems and their physical operations are abstracted as quantum gates, even though these 2 systems have different physical properties. (Middle) SC qubits are SC circuits placed inside a long, metal transmission line. The apparatus requires a dilution fridge of temperature near absolute zero. The orange standing waves are oscillations in the transmission line, which are  driven by external microwave pulses and used to control the qubit states. (Right) Trapped ion qubits  are confined in the potential of  cylindrical electrodes. Modulated laser beam can provide elementary quantum operations for trapped ion qubits. The apparatus is usually contained inside a vacuum chamber of pressure around $10^{-8}Pa$. The two systems require different high-level optimizations for better efficiency due to their distinct physical features. }
    \label{fig:hardware}
\end{figure*}
Currently, major quantum programming environments, including Qiskit \cite{Qiskit} by IBM, Cirq \cite{Cirq} by Google, PyQuil \cite{pyquil} by Rigetti and strawberry fields \cite{strawberry} by Xanadu, follow the quantum circuit model. These programming environments support users in configuring, compiling and running their quantum programs  in an automated workflow and roughly follow a layered approach as illustrated in \cref{fig:layers}. In these environments, the compilation stack is divided into layers of subroutines that are built upon the abstraction provided by the next layer. This design philosophy is similar to that of its classical counterpart, which has slowly converged to this layered approach over many years to manage the increasing complexity that comes with the exponentially growing hardware resources. In each layer, burdensome hardware details are well encapsulated and hidden behind a clean interface, which offers a well-defined, manageable optimization task to solve. Thus, this layered approach provides great portability and modularity. For example, the Qiskit compiler supports both the superconducting QIP platform and the trapped ion QIP platform as the backend (\cref{fig:hardware}). In the Qiskit programming environment, these two backends share a unified, hardware-agnostic programming frontend even though the hardware characteristics and the qubit control methods of the two platforms are rather different. Superconducting (SC) qubits are macroscopic LC circuits  placed inside dilution fridges of temperature near absolute zero. These qubits can be regarded as artificial atoms and are protected by a metal transmission line from environmental noise. For SC QIP platforms, qubit control is achieved through sending microwave pulses into the transmission line that surrounds the LC circuits to change the qubit state and those operations are  usually done within several hundreds of nanoseconds. On the other hand, trapped ion qubits are ions confined in the potential of electrodes in vacuum chambers. Trapped ion qubits have a much longer coherence ($>1$ second) and quantum operations are performed by shinning modulated laser beam. The quantum gates are also much slower than that of SC qubits but the qubit connectivity (for 2-qubit gates) are much better. In Qiskit, the  hardware characteristics of the two QIP platforms are abstracted away in the quantum circuit model so that the higher level programming environment can work with both backends.

However, the abstractions introduced in the layered approach of current QC stacks restrict\ opportunities for cross-layer optimizations. For example, without accessing the lower level noise information, the compiler might not be able to properly optimize gate scheduling and qubit mapping with regard to the final fidelity. For near-term quantum computing, maximal utilization of the scarce quantum resources and reconciling quantum algorithms with noisy devices is of more importance than to manage complexity of the classical control system.  In this review, we propose a shift of the quantum computing stack towards a more vertical integrated architecture. We point out that breaking the abstraction layers in the stack by exposing enough lower level details could substantially improve the quantum efficiency. This claim is not that surprising --- there are many supporting examples from the classical computing world such as the emergence of application specific architectures like the GPU and the TPU. However, this view is often overlooked in the software/hardware design in quantum computing. 

We examine this methodology by looking at several previous works along this line. We first review two compilation optimizations that break the ISA abstraction by exposing pulse level information (\cref{sec:pulse}) and noise information (\cref{sec:noise}), respectively. Then, we discuss an information processing scheme that improves general circuit latency by exposing the third energy level of the underlying physical space, $i.e.$, breaking the qubit abstraction using qutrits (\cref{sec:qutrit}). Then, we discuss the Gottesman-Kitaev-Preskill (GKP) qubit encoding in a Quantum Harmonic Oscillator (QHO) that exposes error information in the form of small shifts in the phase space to assist the upper level error mitigation/correction procedure (\cref{sec:gkp}).

At last, we envision several future directions that could further explore the idea of breaking abstractions and assist the realization of the first quantum computers for real-world applications.

\section{Breaking the ISA Abstraction using Pulse-Level Compilation}
\label{sec:pulse}

In this section, we describe a quantum compilation methodology proposed in \cite{ShiASPLOS19,partialPranav} that achieves an average of 5X speedup in terms of generated circuit latency, by employing the idea of breaking the ISA abstraction and compiling directly to control pulses. 

\subsection{Quantum compilation}

Since the early days of quantum computing, quantum compilation has been recognized  as one of the central tasks in realizing practical quantum computation. Quantum compilation was first defined as the problem of synthesizing quantum circuits for a given unitary matrix. The celebrated Solovay-Kitaev theorem \cite{kitaev94} states that such synthesis is always possible if a universal set of quantum gates is given. Now the term of quantum compilation is used more broadly and almost all stages in \cref{fig:layers} can be viewed as part of the quantum compilation process.

There are many indications that  current quantum compilation stack (\cref{fig:layers}) is highly-inefficient. First, current circuit synthesis algorithms are far from saturating (or being closed to) the asymptotic lower bound in the general case \cite{Nielsen, kitaev94}.  Also, the formulated circuit synthesis problem is based on the fundamental abstraction of quantum ISA (\cref{lgate}) and largely discussed in a hardware-agnostic settings in previous work but the underlying physical operations cannot be directly described by the logical level ISA (as shown in \cref{fig:hardware}). The translation from the logical ISA to the operations directly supported by the hardware is typically done in an {\it ad-hoc} way. Thus, there is a mismatch between the expressive logical gates and the set of instructions that can be efficiently implemented on a real system. This mismatch significantly limits the efficiency of the current quantum computing stack thus the underlying quantum devices’ computing ability and wastes precious quantum coherence. While improving the computing efficiency is always valuable, improving quantum computing efficiency is do-or-die: computation has to finish before qubit decoherence or the results will be worthless. Thus, improving the compilation process is one of the most, if not the most, crucial goals in near-term QC system design.

By identifying this mismatch and the fundamental limitation in the ISA abstraction, in \cite{ShiASPLOS19, partialPranav}, we proposed a quantum compilation technique that optimizes across existing abstraction barriers to greatly reduce latency while still being practical for large numbers of qubits. Specifically, rather than limiting the compiler to use 1- and 2-qubit quantum instructions, our framework aggregates the instructions in the logical ISA into a customized set of instructions that correspond to optimized control pulses. We compare our methodology to the standard  compilation workflow on several promising NISQ quantum applications and conclude that our compilation methodology has an average speedup of $5 \times$ with a maximum speedup of  $10\times$. We use the rest of this section to introduce this compilation methodology, starting with defining some basic concepts.

\subsection{Quantum ISA}\label{lgate}
In the quantum computing stack, a restricted set of 1- and 2-qubit quantum instructions are provided for describing the high-level quantum algorithms, analogous to the Instruction Set Architecture (ISA) abstraction in classical computing. In this paper, we call this instruction set the {\it logical ISA}. The 1-qubit gates in the logical ISA include the Pauli gates, $P=\{X, Y, Z\}$. It also includes the Hadamard $H$ gate, whose symbol in the circuit model is given as an example in \cref{fig:hardware} on the left column. The typical 2-qubit instruction in the logical instruction set is the Controlled-NOT (CNOT) gate, which flips the state of the target qubit based on the state of the control qubit.

However, usually quantum computing devices does not directly support the logical ISA. Based on the system characteristics, we can define the  {\it physical ISA} that can be directly mapped to the underlying control signals. For example, superconducting devices typically has Cross-Resonance (CR) gate or $i$SWAP gate as their intrinsic 2-qubit instruction, whereas for trapped-ion devices the intrinsic 2-qubit instruction can be the Mølmer–Sørensen gate or the controlled phase gate.

\subsection{Quantum Control}

As shown in \cref{fig:hardware} and discussed in the last subsection, underlying physical operations in the hardware such as microwave control pulses and modulated laser beam are abstracted as quantum instructions. A quantum instruction are simply as pre-fined control pulse sequences. 

The underlying evolution of the quantum system is continuous and so are the control signals. The continuous control signals offer much richer and flexible controllability than the quantum ISA. The control pulses can drive the quantum computing hardware to a desired quantum states by varying a system-dependent and time-dependent quantity called the Hamiltonian. The Hamiltonian of a system determines the evolution path of the quantum states. The ability to engineer real-time system Hamiltonian  allows us to navigate the quantum system to the quantum state of interest through generating accurate control signals. Thus, quantum computation can be done by constructing a quantum system in which the system Hamiltonian evolves in a way that aligns with a quantum computing task, producing the computational result with high probability upon final measurement of the qubits. In general, the path to a final quantum state is not unique and finding the optimal evolution path is a very important but challenging problem \cite{Nelson2017, CISC2007, Glaser2015}.

\subsection{The Mismatch Between ISA and Control}
\label{mismatch}
Being hardware-agnostic, the quantum operation sequences composed by logical ISA have limited freedom in terms of controllability and usually will not be mapped to the optimal evolution path of the underlying quantum system, thus there is a mismatch between the ISA and  low-level quantum control. With two simple examples, we demonstrate this mismatch.
\begin{itemize}
    \item We can consider the instruction sequence consists of a CNOT gate followed by a $X$ gate on the control bit. In current compilation workflow, these two logical gates will be further decomposed into the physical ISA and be executed sequentially. However, on SC QIP platforms, the microwave  pulses that implement these two instructions could in fact be applied simultaneously (because of their commutativity). Even the commutativity can be captured by the ISA abstraction, in the current compilation workflow, the compiled control signals are sub-optimal.
        \item 
SWAP gate is an important quantum instruction for circuit mapping.   The SWAP operation is usually decomposed as  3 Controlled-NOT (CNOT) operations, as realized in the circuit below. This decomposition could be thought of the implementation of in-place memory SWAPs with three alternating XORs for classical computation. However, for systems like quantum dots \cite{quantumdot}, the SWAP operation is directly supported by applying particular constant control signals for a certain period of time. In this case, this decomposition of SWAP into three CNOTs introduces substantial overhead.

 \begin{figure}[h]
      \centering
\hspace*{-0.1cm}\includegraphics[scale=0.25]{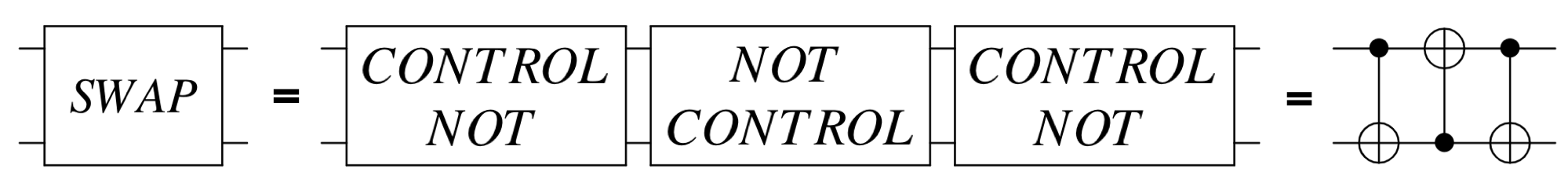} 
\label{swap_decomposition}
\end{figure}
\end{itemize}
In experimental physics settings, equivalences from simple gate sequences to control pulses can be hand optimized \cite{schuch2003}. However, when circuits become larger and more complicated, this kind of hand optimization become less efficient and the standard decomposition becomes less favorable,  motivating a shift toward numerical optimization methods that are not limited by the ISA abstraction.

\subsection{Quantum Optimal Control}

 Quantum Optimal Control (QOC) theory provides an alternative in terms of finding the optimal evolution path for the quantum compilation tasks. Quantum optimal control algorithms typically perform analytical or numerical methods for this optimization, among which, gradient ascent methods, such as the GRadient Ascent Pulse Engineering (GRAPE) \cite{grape1, grape2} algorithm, are widely used. The basic idea of GRAPE is as follows: for optimizing the control signals of $M$ parameters ($u_1,\ldots,u_M$) for a target quantum state, in every iteration, GRAPE minimizes the deviation of the system evolution by calculating the gradient of the final fidelity with respect to the $M$ control parameters in the $M$ dimensional space. Then GRAPE will update the paramters in the direction of the gradient with adaptive step size \cite{grape1, grape2, Nelson2017}. With a large number of  iterations, the optimized control signals are expected to converge and find optimized pulses.

 In \cite{ShiASPLOS19}, we utilize GRAPE to optimize our aggregated instructions that are customized for each quantum circuit as opposed to selecting instructions from a pre-defined pulse sequences. However, one disadvantage of numerical methods like GRAPE is that the running time and memory use grow exponentially with the size of the quantum system for optimization. In our work, we are able to use GRAPE for optimizing quantum systems of up to ~10 qubits with the GPU accelerated optimal control unit \cite{Nelson2017}. As shown in our result, the limit of 10 qubits does not put restrictions on the result of our compilation methodology.
 
 \subsection{Pulse-Level Optimization: A Motivating Example}\label{qaoa}
\begin{figure}[t]
\centering
\includegraphics[width=0.46\textwidth]{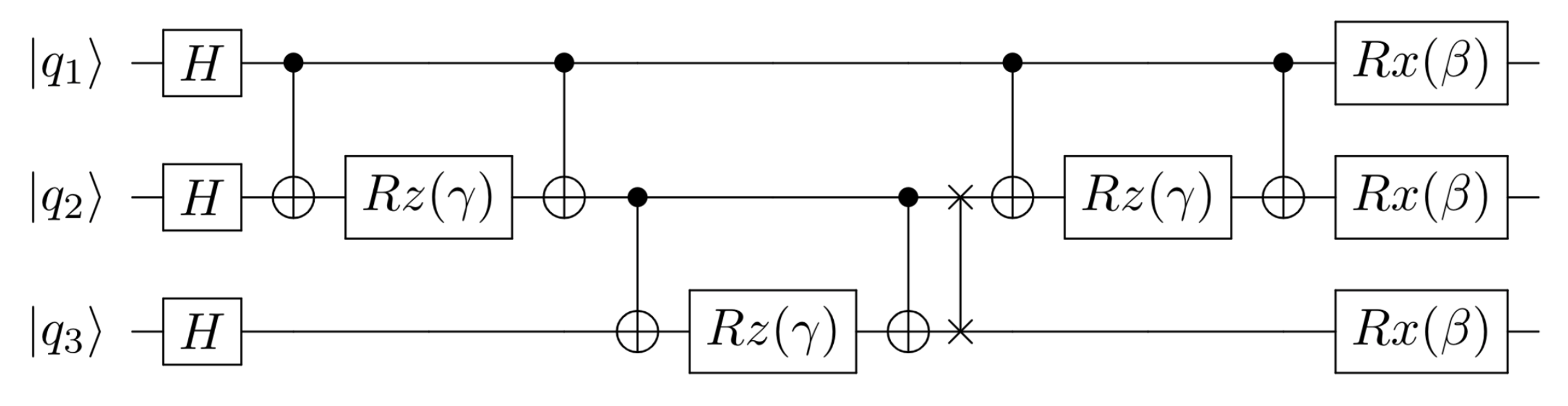}
\newline
\hspace*{0.3cm}(a)
\includegraphics[width=0.46\textwidth]{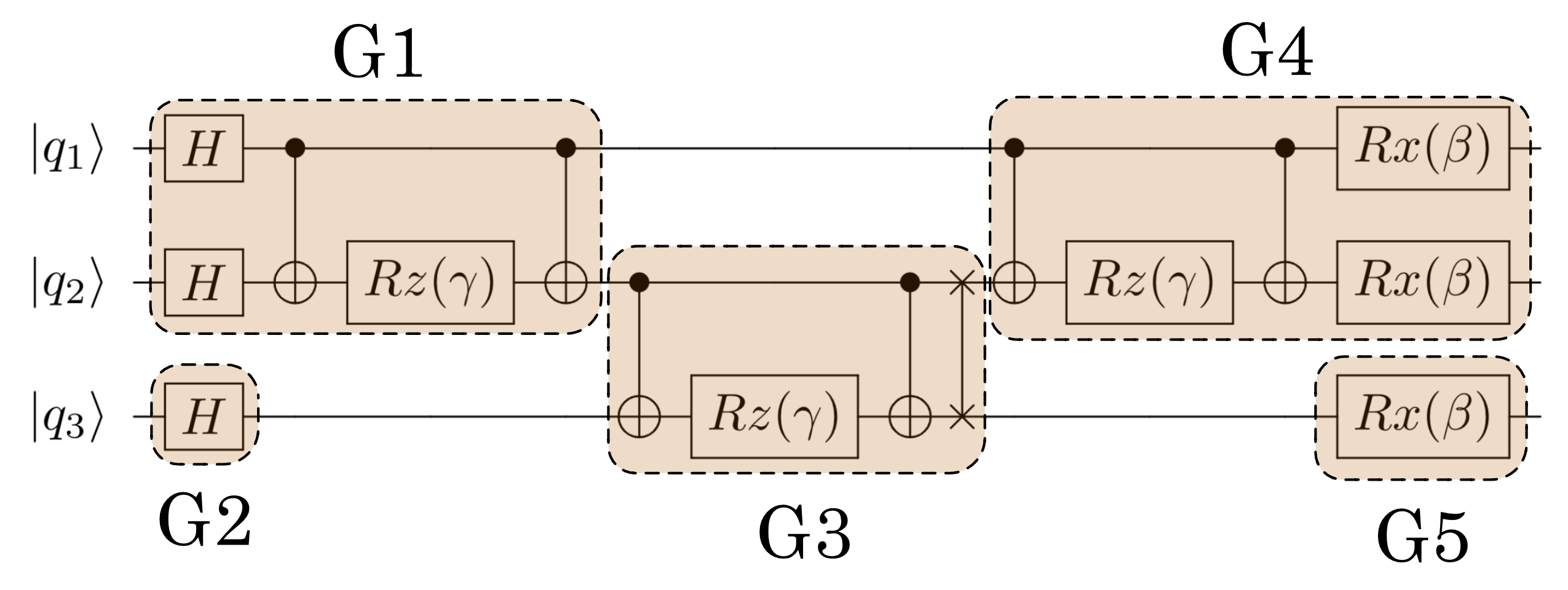}
\vspace*{0.25cm}
\hspace*{0.3cm}(b)
\includegraphics[width=0.45\textwidth]{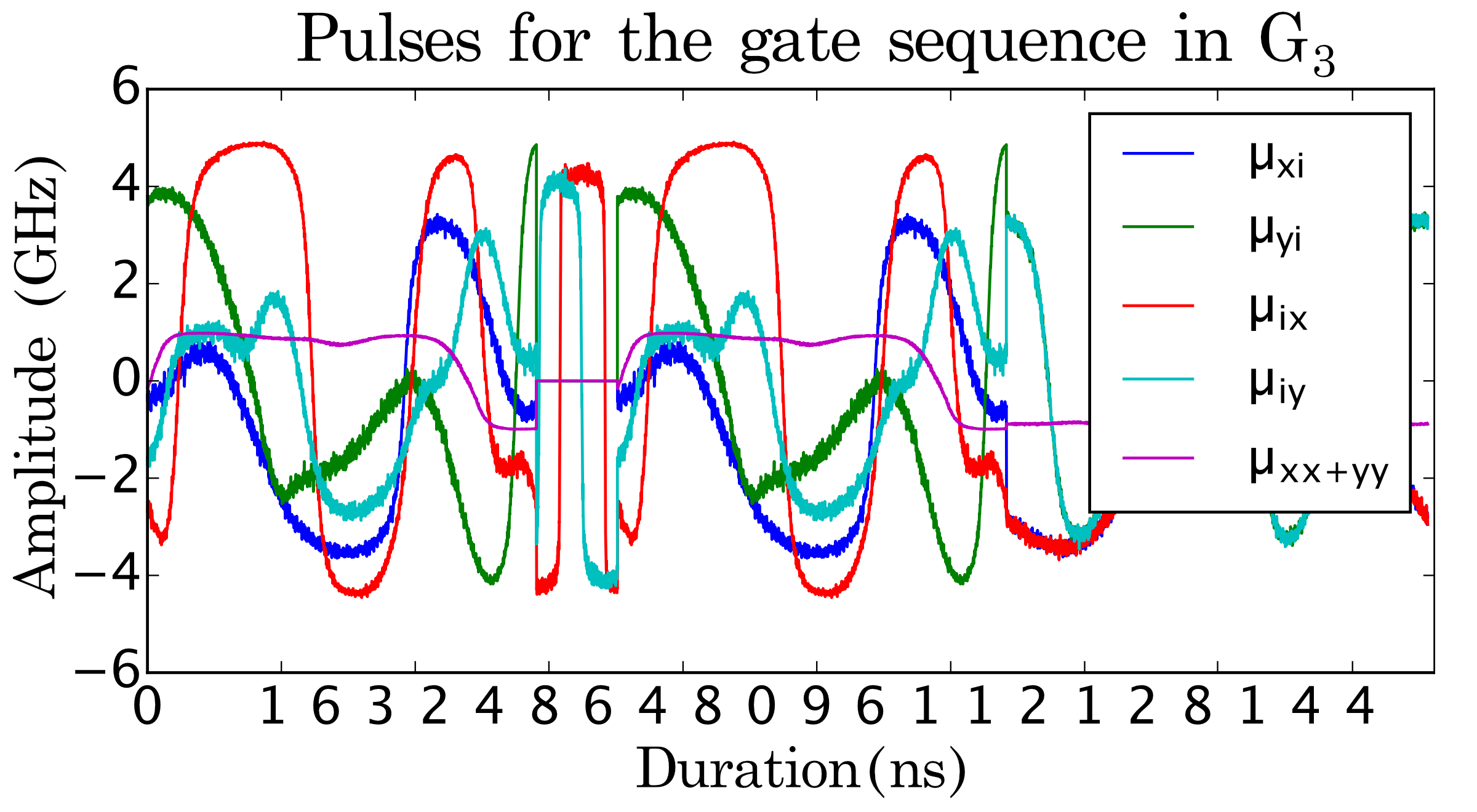}
\vspace*{0.25cm}
\hspace*{0.3cm}(c)
\includegraphics[width=0.45\textwidth]{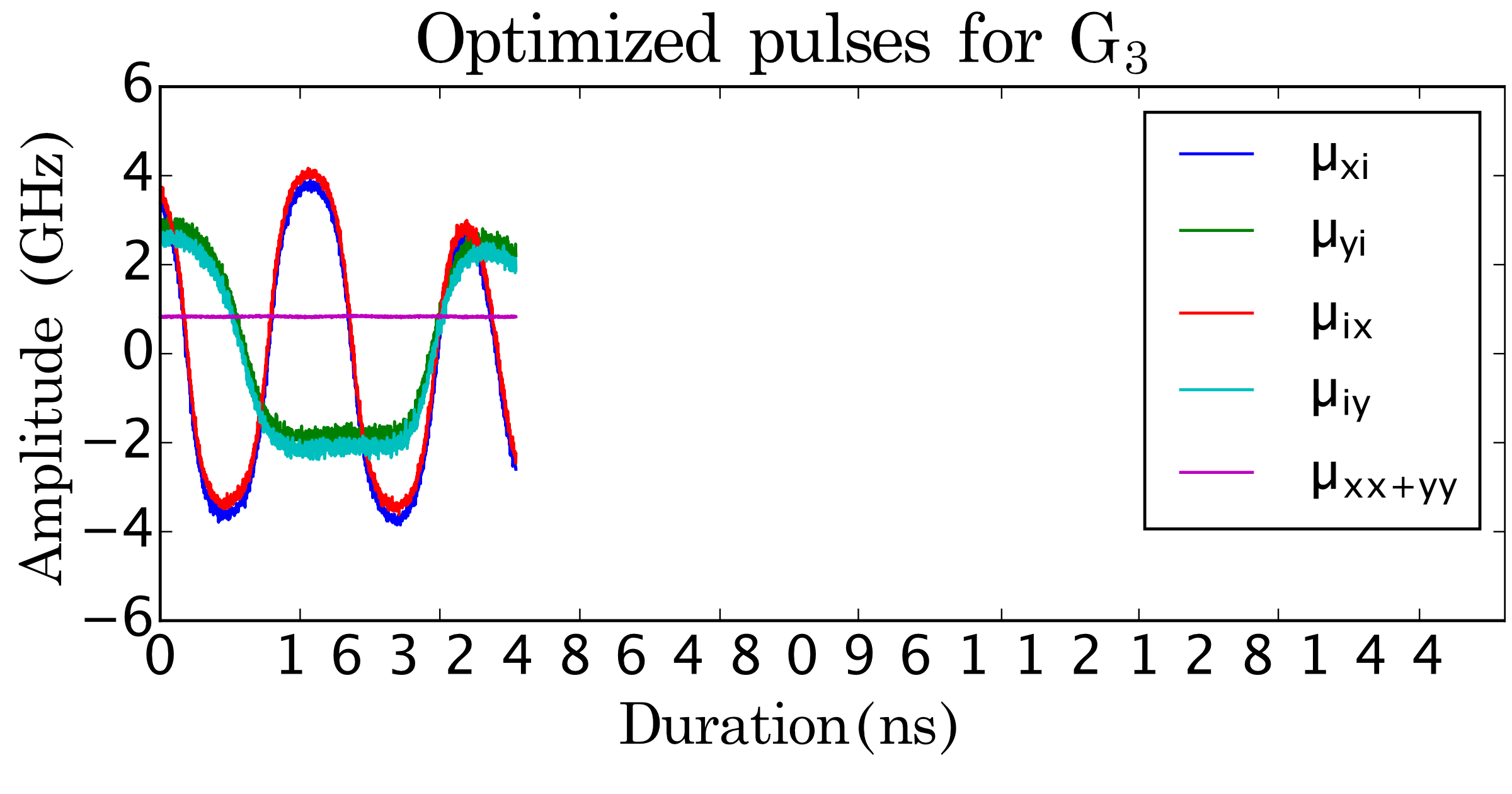}
\hspace*{0.3cm}(d)
\caption{Example of a QAOA circuit. (a) The QAOA circuit with the logical ISA. (b) The QAOA circuit with aggregated instructions. (c) Generated control pulses for $G_3$ in the ISA-based compilation. (d) Control pulses for $G_3$ from aggregated instructions based compilation. Each curve is the amplitude of a relevant control signal. The pulse sequences in (d) provides a $3\times$ speedup comparing to the pulse sequences in (c). Pulse sequences reprinted with permission from \cite{ShiASPLOS19}}
\label{demo_qaoa}
\end{figure}

Next, we will illustrate the workflow of our  compilation methodology with a circuit instance of the Quantum Approximate Optimization Algorithm (QAOA) for solving the MAXCUT problem on the triangle graph (\cref{demo_qaoa}). \footnote{The angle parameters $\gamma$ and $\beta$ can be determined by variational methods \cite{Mcclean2016} and are set to $5.67$ and $1.26$.}
This QAOA circuit with logical ISA (or variants of it up to single qubit gates) can be reproduced by most existing quantum  compilers. This instance of the QAOA circuit is generated by the ScaffCC compiler, as shown in \cref{demo_qaoa} (a). We assume this circuit is executed on a  superconducting architecture with 1D nearest neighbor qubit connectivity. A SWAP instruction is inserted in the circuit to satisfy the linear qubit connectivity constraints. 
\begin{figure*}[thpb]
      \centering
 \hspace*{-0.3cm}\includegraphics[width=0.88\textwidth]{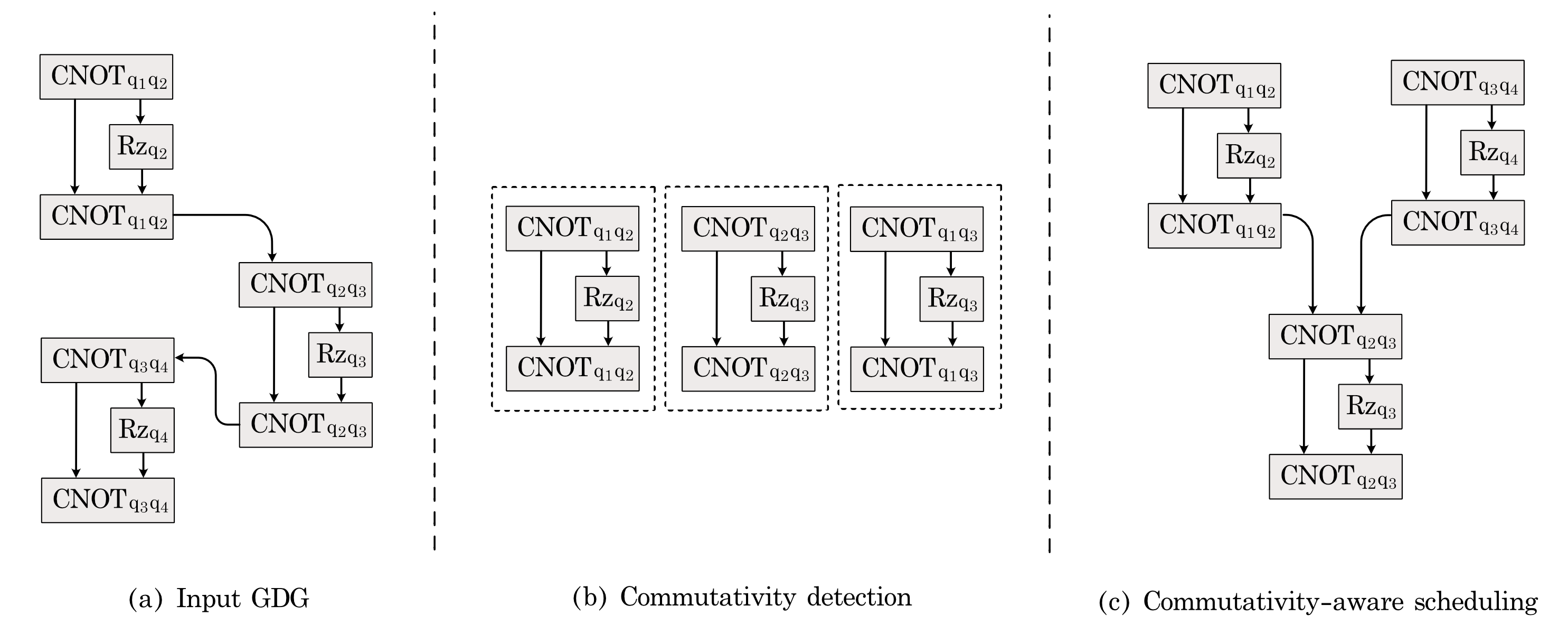}
   \caption{The example in \cref{demo_comm} in the form of gate dependency graph. } 
      \label{gdg}
   \end{figure*}

 On the other hand, our compiler generates the aggregated instruction set $G_1-G_5$ as illustrated in \cref{demo_qaoa} (b) automatically, and uses GRAPE to produce highly-optimized pulse sequences for each aggregated instruction. 
In this minimal circuit instance, our  compilation method reduces the total execution time of the circuit by about $2.97\times$ comparing to compilation with restricted ISA. \cref{demo_qaoa} (c) and (d) show the generated pulses for $G_3$ with ISA-based compilation and with our  aggregated instruction based, pulse-level optimized compilation.

\subsection{Optimized Pulse-Level Compilation Using Gate Aggregation: the workflow}
\begin{figure}[t]
\centering
\includegraphics[width=0.38\textwidth]{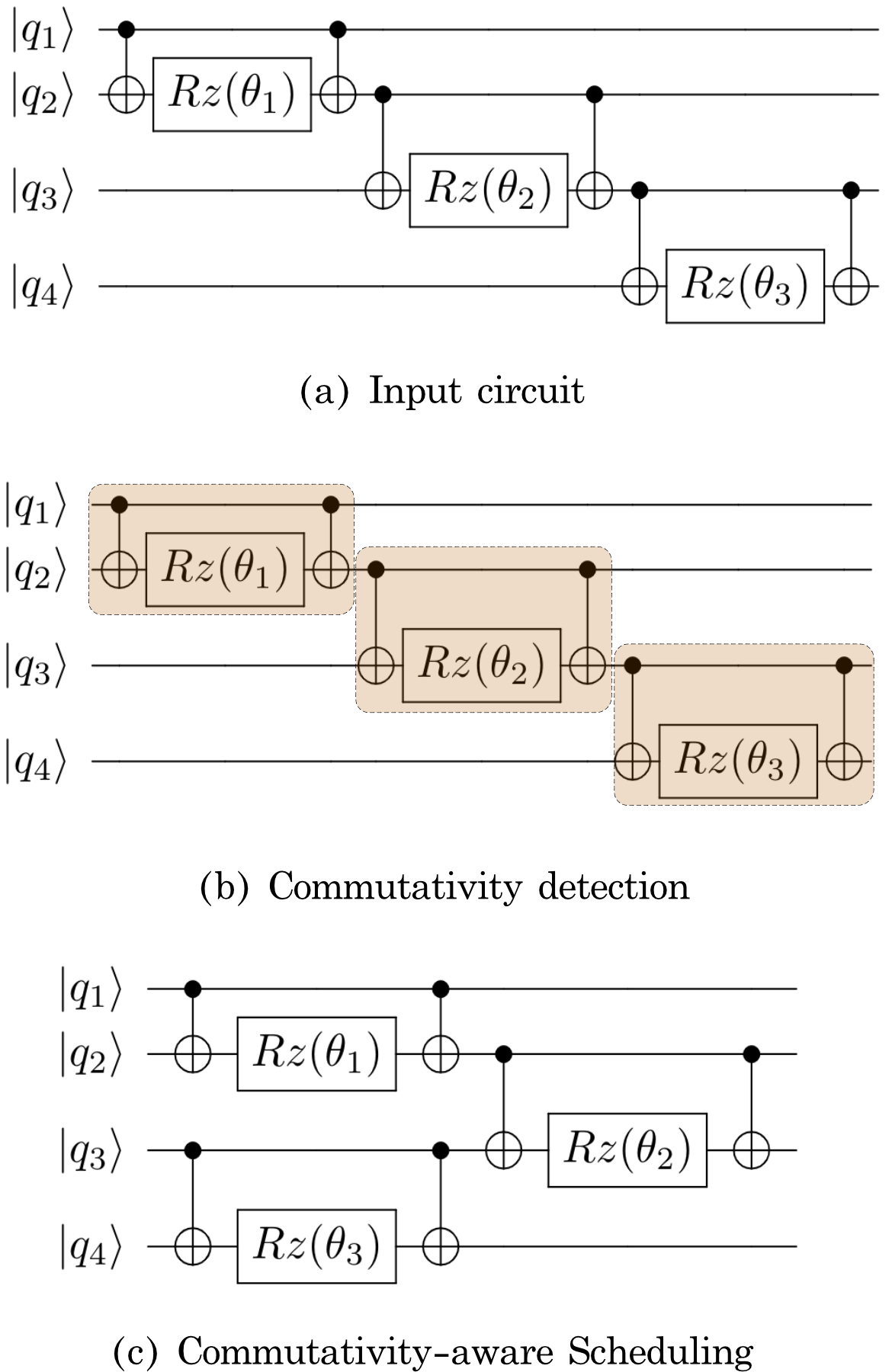}

\caption{An example of commutativity-aware logical scheduling. With commutativity detected, the circuit depth can be shortened.}
\label{demo_comm}
\end{figure}

   Now we give a systematic view of the workflow of our compiler. First, at the program level, our compiler performs module flattening and loop unrolling to produce the quantum assembly (QASM), which represents a schedule of the logical operations. Next, the compiler enters the commutativity detection phase.
   Different  from the ISA-based approach, in this phase, our compilation process  converts the QASM code to a more flexible logical schedule that explores the commutativity between instructions.  To further explore the commutativity in the schedule, the compiler aggregates instructions in the schedule to produce a new logical schedule with instructions that represents diagonal matrices (and are of high commutativity).  Then the compiler enters the scheduling and mapping phase. Because of commutativity awareness, our compiler can generate a much more efficient logical schedule by re-arranging the aggregated instructions with high commutativity. The logical schedule is then converted to a physical schedule after the qubit mapping stage. Then the compiler generates the final aggregated instructions for pulse optimization and use GRAPE for producing the corresponding control pulses. The goal of the final aggregation is to find the optimal instruction set that produces the lowest-latency control pulses while preserving the parallelism in the circuit  aggregations that are small as much as possible. Finally, our compiler  outputs an optimized physical schedule along with the corresponding optimized control pulses.
 \cref{gdg} shows the Gate Dependency Graph (GDG) of the QAOA circuit in Figure \ref{demo_comm} in different compilation stages. Next, we walk through the compilation backend with this example, starting from the commutativity detection phase. 
\subsubsection{Commutativity detection:} \label{ps_ai}
In the commutativity detection phase, the false dependence between commutative instructions are removed and the GDG is re-structureed. This is because if a pair of gates commutes, the gates can be scheduled in either order. Also, it can be further noticed that, in many NISQ quantum algorithms, it is ubiquitous that for instructions \emph{within} an instruction block to not commute, but for the full instruction block to commute with each other \cite{Farhi2014, qchemistry}. As an example, in Figure \ref{demo_comm}, the CNOT-Rz-CNOT instruction block commute with each other because these blocks correspond to diagonal unitary matrices. However, each individual instruction in these circuit blocks does not commute. Thus , after aggregating these instructions, the compiler is able to schedule new aggregated instructions in any order, which is impossible before. This commutativity detection procedure opens up opportunities for more efficient scheduling. 

\subsubsection{Scheduling and mapping:}\label{cschedule}
\textbf{Commutativity-aware logical scheduling (CLS)} In our scheduling phase, our logical scheduling algorithm is able to fully utilize the detected commutativity in the last compilation phase. The CLS iteratively schedules the available intructions on each qubits.  At each iteration, the CLS draws instruction candidates that can be executed in the earlist time step to schedule.

\textbf{Qubit mapping} In this phase of the compilation, the compiler transform the circuit to a form that respect the topological constraints of hardware connectivity \cite{Maslov2008QuantumCP}. To conform to the device topology, the logical instructions are processed in two steps. First, we place frequently interacting qubits near each other by bisecting the qubit interaction graph along a cut with few crossing edges, computed by the METIS graph partitioning library \cite{METIS}.
Once the initial mapping is generated, two-qubit operations between non-neighboring qubits are prepended with a sequence of SWAP rearrangements that move the control and target qubits to be adjacent.

\subsubsection{Instruction aggregation:}\label{final_ai}
 In this phase, the compiler iterates with the optimal control unit to generate the  final aggregated instructions for the circuit. Then, the optimal control unit optimizes each instruction individually with GRAPE. 

\subsubsection{Physical Execution:}\label{exec}
Finally, the circuit will be scheduled again using the CLS from  \cref{cschedule}, the physical schedules will be sent to the control unit of the underlying quantum hardware and trigger the optimized control pulses at appropriate timing and  the physical execution.

\subsection{Discussion}
In \cite{ShiASPLOS19}, we selected 9 important quantum /classical-quantum hybrid algorithms in the NISQ era as our benchmarks. Across all 9 benchmarks, our compilation scheme achieves a geometric mean of 5.07× pulse time reduction comparing to the standard gate-based compilation. The result in \cite{ShiASPLOS19} indicates that addressing the mismatch between quantum gates and the control pulses by breaking the ISA abstraction can greatly improve the compilation efficiency. Going beyond the ISA-based compilation, this work opens up a door to new QC system designs. 
\section{Breaking the ISA abstraction using Noise-Adaptive Compilation}
\label{sec:noise}
\begin{figure}[t]
\centering
\subfloat[Coherence time (T2)]
{
    \includegraphics[scale=0.5]{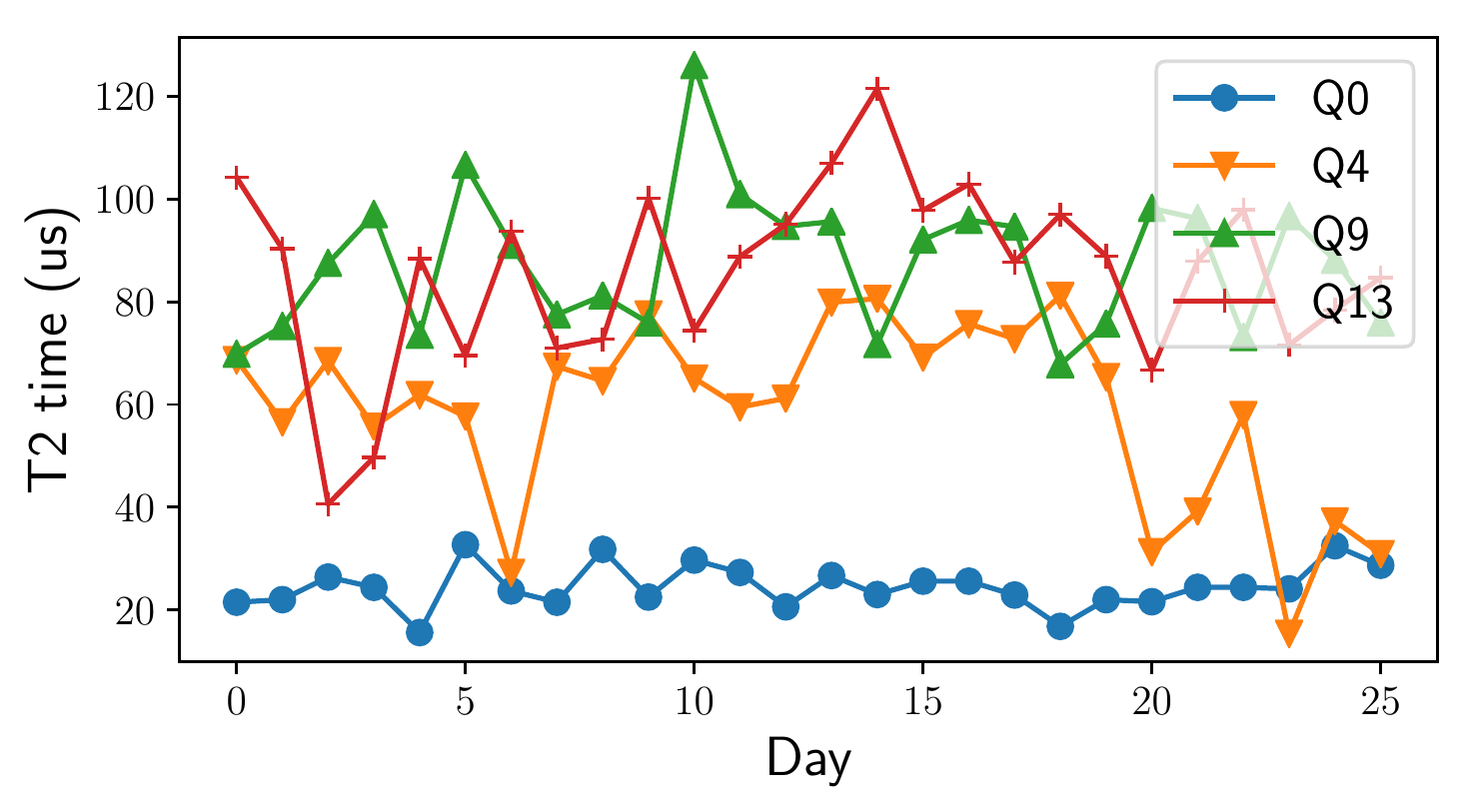}
    \label{fig:ibmqx5_coherence}
}
\vspace{-1em}
\subfloat[CNOT gate error rate]
{
    \includegraphics[scale=.5]{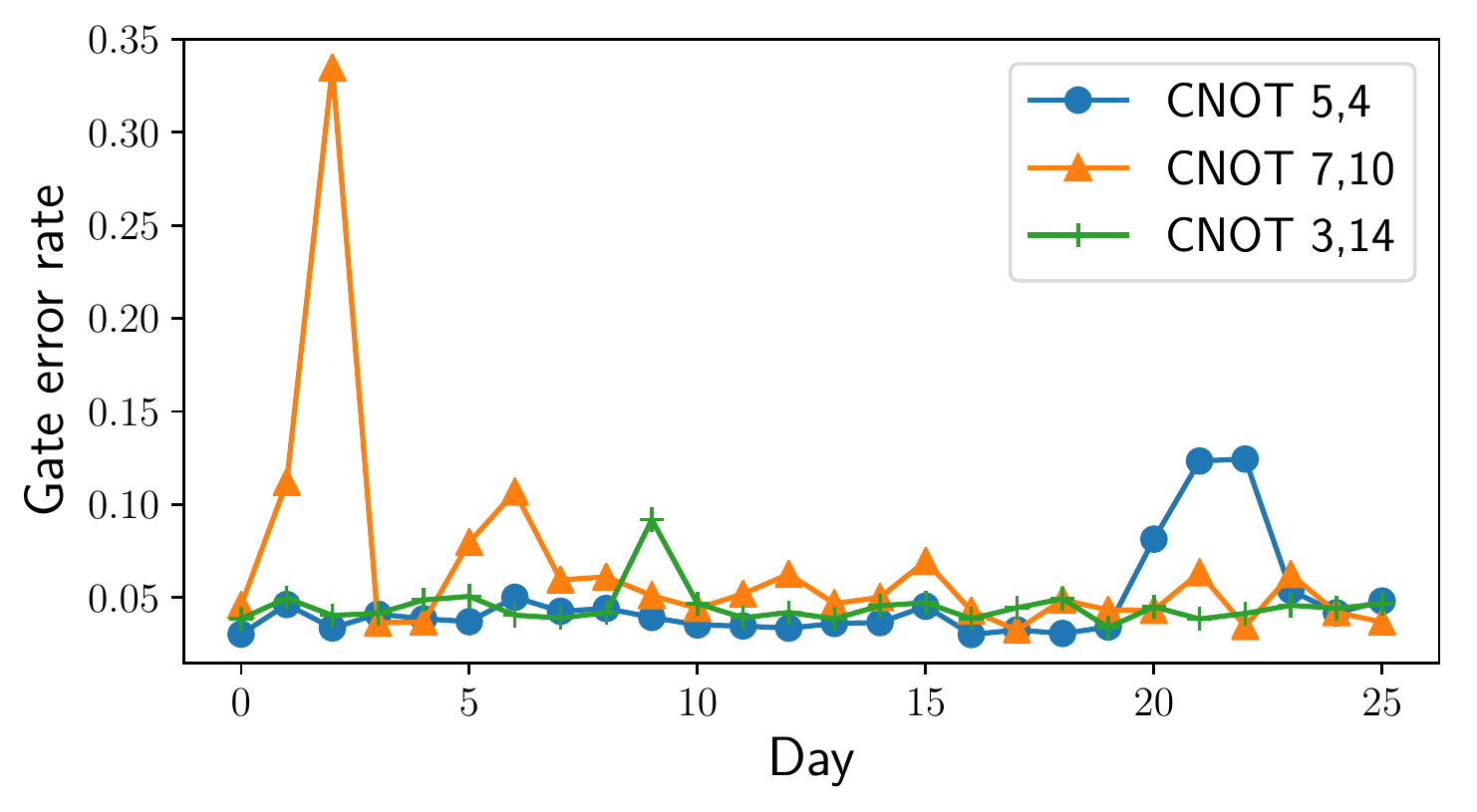}
    \label{fig:ibmqx5_cx_error}
}
\caption{Daily variations in qubit coherence time (larger is better) and gate error rates (lower is better) for selected qubits and gates in IBM's 16-qubit system. The most or least reliable system elements change across days.}
\label{fig:hw_var2}
\end{figure}
\begin{figure*}
\centering
\subfloat[Bernstein-Vazirani Intermediate Representation]
{
    \includegraphics[scale=0.35]{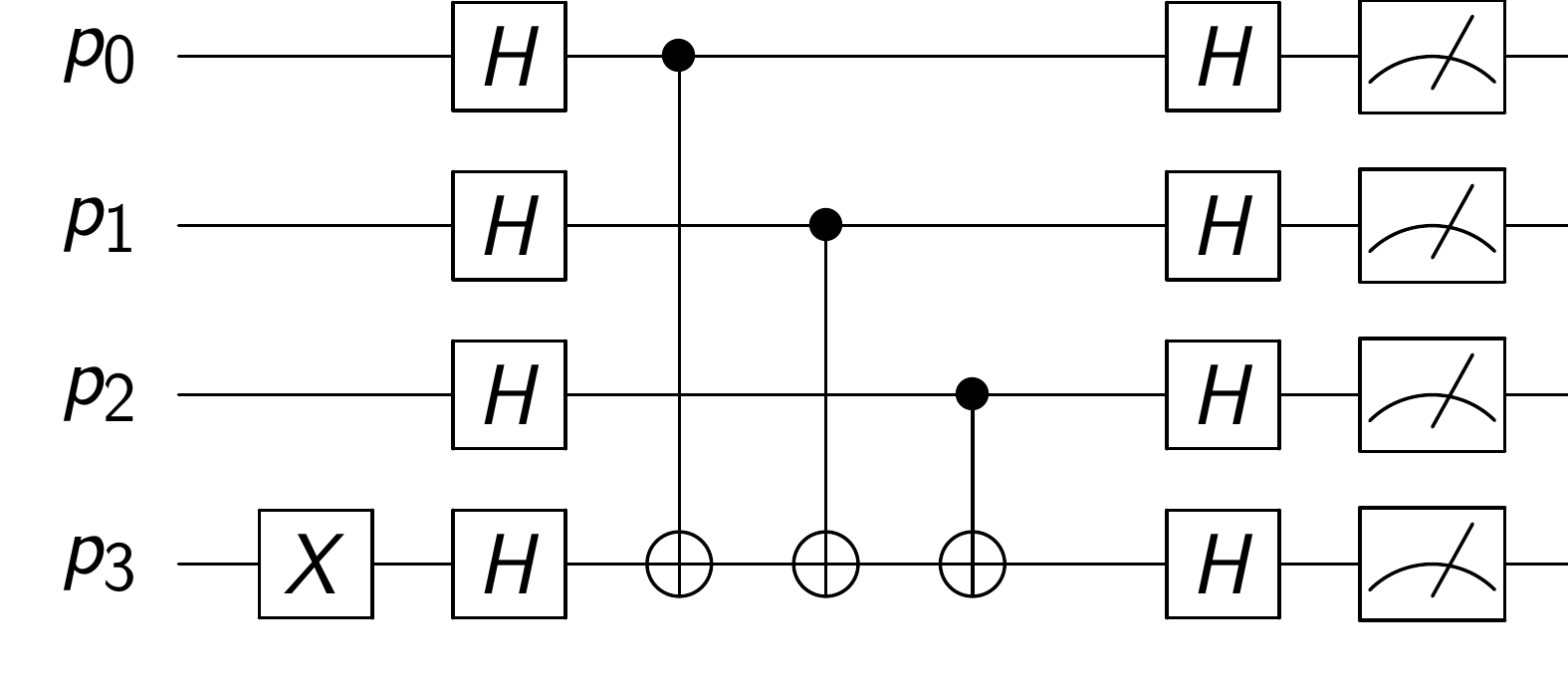}
    \label{fig:bv4_code}
}
\qquad
\subfloat[Layout of qubits in IBMQ16 and a naive mapping for BV4.]
{
    \includegraphics[scale=.45]{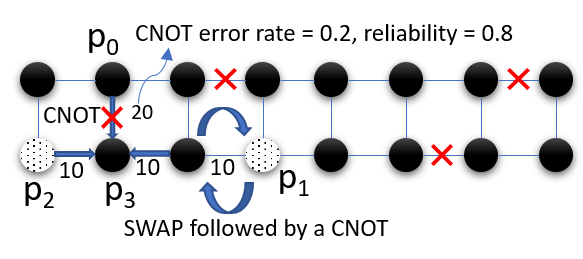}
    \label{fig:bv4_random}
}
\qquad
\subfloat[Optimized mapping for BV4]
{
    \includegraphics[scale=.4]{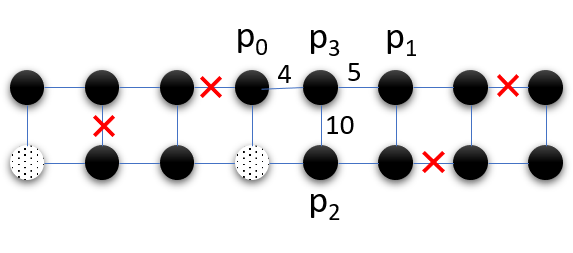}
    \label{fig:bv4_goodmap}
}
\label{fig:bv_mappings}
\caption{Figure (a) shows the intermediate representation of the Bernstein-Vazirani algorithm (BV4). Each horizontal line represents a program qubit. X and H are single qubit gates. The Controlled NOT (CNOT) gates from each qubit $p_{0,1,2}$ to $p_3$ are marked by vertical lines with XOR connectors. The readout operation is indicated by the meter. Figure (b) shows the qubit layout in IBMQ16, a naive mapping of BV4 onto this system. The black circles denote qubits and the edges indicate hardware CNOT gates. The edges are labelled with CNOT gate error ($\cross10^{-2}$). The hatched qubits and crossed gates are unreliable. In this mapping, a SWAP operation is required to perform the CNOT between $p_1$ and $p_3$ and error-prone operations are used. Figure (c) shows a mapping where qubit movement is not required and unreliable qubits and gates are avoided.}
\end{figure*}
In recent years, QC compute stacks have been developed using abstractions inspired from classical computing. 
The instruction set architecture (ISA) is a fundamental abstraction which defines the interface between the hardware and software. The ISA abstraction allows software to execute correctly on any hardware which implements the interface. This enables application portability and decouples hardware and software development.

For QC systems, the hardware-software interface is typically defined as a set of legal instructions and the connectivity topology of the qubits \cite{openqasm1, pyquil, koen_bertels1, koen_bertels2, koen_bertels3},  --- it does not include information about qubit quality, gate fidelity or micro-operations used to implement the ISA instructions. 
While technology independent abstractions are desirable in the long run, our work \cite{pmasplos, triq} reveals that such abstractions are detrimental to program correctness in NISQ quantum computers. By exposing microarchitectural details to software and using intelligent compilation techniques, we show that program reliability can be improved significantly.

\subsection{Noise Characteristics of QC Systems}
QC systems have spatial and temporal variations in noise due to manufacturing imperfections, imprecise qubit control and external interference. To motivate the necessity for breaking the ISA abstraction barrier, we present real-system statistics of hardware noise in systems from three leading QC vendors -- IBM, Rigetti and University of Maryland. IBM and Rigetti systems use superconducting qubits \cite{ibm_cr_gate, rigetti_transmons} and University of Maryland (UMD) uses trapped ion qubits \cite{trappedion3}. The gates in these systems are periodically calibrated and their error rates are measured.

Figure \ref{fig:hw_var2} shows the coherence times and two-qubit gate error rates in IBM's 16-qubit system (\ibmnamefull). From daily calibration logs we find that, the average qubit coherence time is 40 microseconds, two-qubit gate error rate is 7\%, readout error rate is 4\% and single qubit error rate is 0.2\%. The two-qubit and readout errors are the dominant noise sources and vary up to 9X across gates and calibration cycles. Rigetti's systems also exhibit error rates and variations of comparable magnitude. These noise variations in superconducting systems emerge from material defects due to lithographic manufacturing, and are expected in future systems also \cite{superconducting_stability, mit_review}. 

Trapped ion systems also have noise fluctuations even though the individual qubits are identical and defect-free. On a 5-qubit trapped ion system from UMD, we observed up to $3$x variation in the two-qubit gate error rates because of fundamental challenges in qubit control using lasers and their sensitivity to motional mode drifts from temperature fluctuations. 

We found that these microarchitectural noise variations dramatically influence program correctness. When a program is executed on a noisy QC system, the results may be corrupted by gate errors, decoherence or readout errors on the hardware qubits used for execution. Therefore, it is crucial to select the most reliable hardware qubits to improve the {\em success rate} of the program (the likelihood of correct execution). The success rate is determined by executing a program multiple times and measuring the fraction of runs that produce the correct output. High success rate is important to ensure that the program execution is not dominated by noise.

\subsection{Noise-Adaptive Compilation: Key Ideas}

\begin{figure}
    \centering
\includegraphics[scale=0.52]{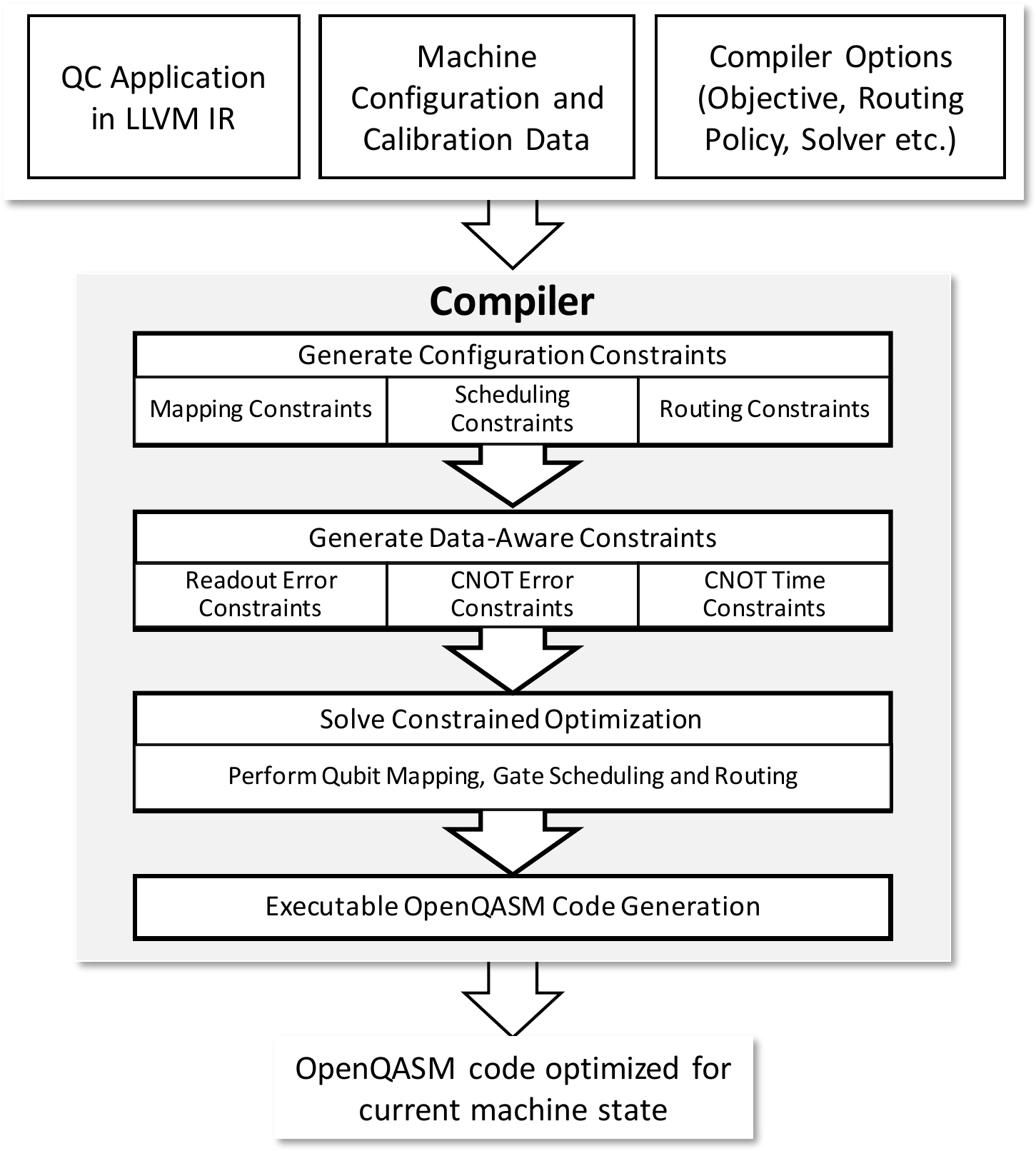}
    \caption{Noise-adaptive compilation using SMT Optimization. Inputs are a QC program IR, details about the hardware qubit configuration, and a set of options, such as routing policy and solver options. From these, compiler generates a set of appropriate constraints and uses them to map program qubits to hardware qubits and schedule operations. The output of the optimization is used to generate an executable version of the program.}
    \label{fig:pipeline}
\end{figure}

\begin{figure*}[t]
   \centering
\subfloat[IBM Qiskit]
{
    \includegraphics[scale=.6]{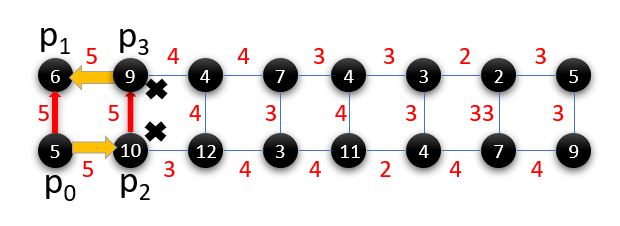}
}
\subfloat[\tSMTstar:Optimize duration without error data]
{
    \includegraphics[scale=.6]{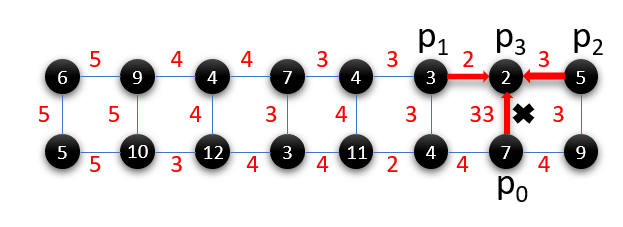}
}\vspace{-4mm}
\subfloat[\rSMTstar ($\omega=1$): Optimize readout reliability]
{
    \includegraphics[scale=.6]{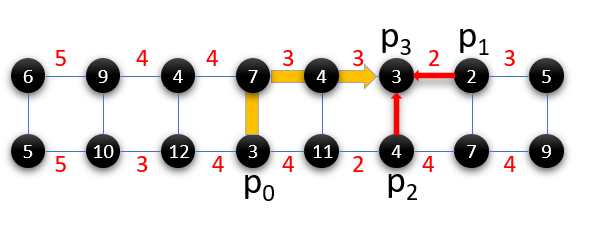}
}
\subfloat[\rSMTstar ($\omega=0.5$): Optimize CNOT$+$readout reliability]
{
    \includegraphics[scale=.6]{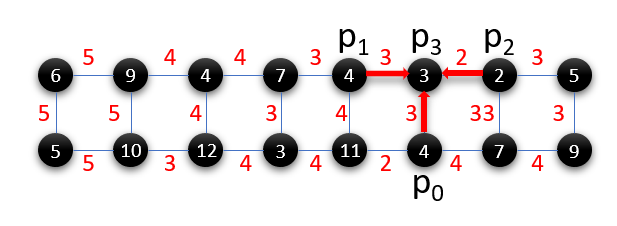}
}
   \caption{For real data/experiment, on IBMQ16, qubit mappings for Qiskit and our compiler with three optimization objectives, varying the type of noise-awareness. The edge labels indicate the CNOT gate error rate ($\cross10^{-2}$), and the node labels indicate the qubit's readout error rate ($\cross10^{-2}$). The thin red arrows indicate CNOT gates. The thick yellow arrows indicate SWAP operations. $\omega$ is a weight factor for readout error terms in the \rSMTstar objective. (a) Qiskit finds a mapping which requires SWAP operations and uses hardware qubits with high readout errors  (b), \tSMTstar finds a a mapping which requires no SWAP operations, but it uses an unreliable hardware CNOT between $p_3$ and $p_0$. (c) Program qubits are placed on the best readout qubits, but $p_0$ and $p_3$ communicate using swaps. (d) \rSMTstar finds a mapping which has the best reliability where the best CNOTs and readout qubits are used. It also requires no SWAP operations.}
   \label{fig:bv_all_mappings}
\end{figure*}

\begin{figure}[h]
    \centering
    \includegraphics[scale=0.3]{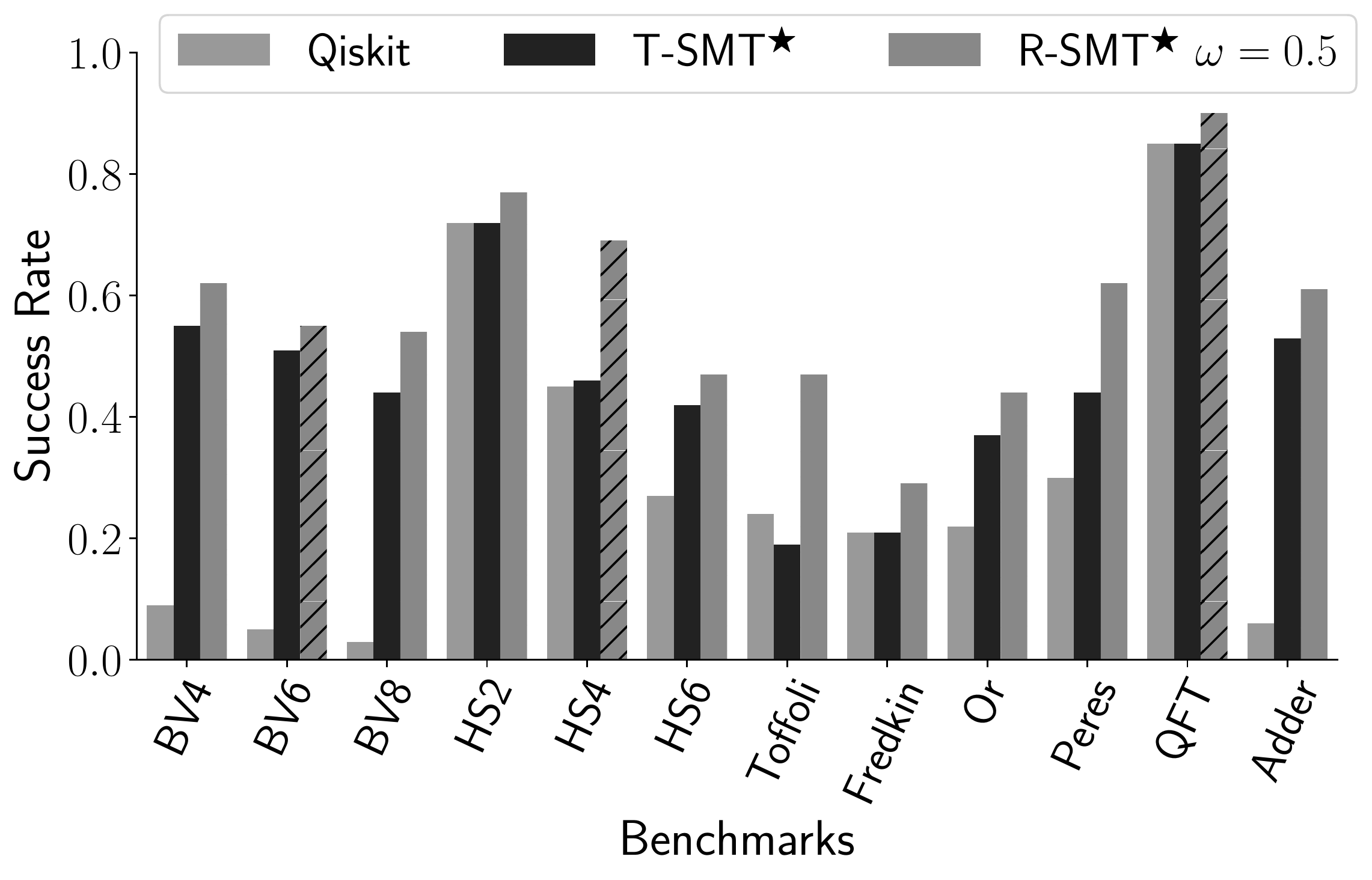}
    \caption{Measured success rate of \rSMTstar compared to Qiskit and \tSMTstar. (Of 8192 trials per execution, success rate is the percentage that achieve the correct answer in real-system execution.)  $\omega$ is a weight factor for readout error terms in the \rSMTstar objective, $0.5$ is equal weight for CNOT and readout errors. \rSMTstar obtains higher success rate than Qiskit because it adapts the qubit mappings according to dynamic error rates and also avoids unnecessary qubit communication.}
    \label{fig:err_full}
\end{figure}
 Our work breaks the ISA abstraction barrier by developing compiler optimizations which use hardware calibration data. These optimizations boost the success rate a program run by avoiding portions of the machine with poor coherence time and operational error rates. 

We first review the key components in a QC compiler. The input to the compiler is a high-level language program (Scaffold in our framework) and the output is machine executable assembly code. First, the compiler converts the program to an intermediate repreesentation (IR) composed of single and two-qubit gates by decomposing high-level QC operations, unrolling all loops and inlining all functions. Figure \ref{fig:bv4_code} shows an example IR. The qubits in the IR (program qubits) are  {\bf mapped} to distinct qubits in the hardware, typically in a way that reduces qubit communication. Next, gates are {\bf scheduled} while respecting data dependencies. Finally, on hardware with limited connectivity, such as superconducting systems, the compiler {\bf inserts SWAP operations} to enable 2-qubit operations between non-adjacent qubits. 

Figure \ref{fig:bv4_random} and \ref{fig:bv4_goodmap} show two compiler mappings for a 4-qubit program on IBM's 16-qubit system. In the first mapping, the compiler must insert SWAPs to perform the two-qubit gate between $p_{1}$ and $p_3$. Since SWAP operations are composed of three two-qubit gates, they are highly error prone. In contrast, the second mapping requires no SWAPs because the qubits required for the CNOTs are adjacent. While SWAP optimizations can be performed using the device ISA, the second mapping is also noise-optimized i.e., it uses qubits with high coherence time and low operational error rates. By considering microarchitectural noise characteristics, our compiler can determine such noise-optimized mappings to improve the program success rate.

We developed three strategies for noise optimization. First, our compiler maps program qubits onto hardware locations with high reliability, based on the noise data. We choose the initial mapping based on two-qubit and readout error rates because they are the dominant sources of error. Second, to mitigate decoherence errors, all gates are scheduled to finish before the coherence time of the hardware qubits. Third, our compiler optimizes the reliability of SWAP operations by minimizing the number of SWAPs whenever possible and performing SWAPs along reliable hardware paths. 

\subsection{Implementation using SMT Optimization}
Our compiler implements the above strategies by finding the solution to a constrained optimization problem using a Satisfiability Modulo Theory (SMT) solver. The variables and constraints in the optimization encode program information, device topology constraints and noise information. The variables express the choices for program qubit mappings, gate start times and routing paths. The constraints specify that qubit mappings should be distinct, the schedule should respect program dependencies and that routing paths should be non-overlapping. Fig. \ref{fig:pipeline} summarizes the optimization-based compilation pipeline for IBMQ16. 

The objective of our optimization is to maximize the success rate of a program execution. Since the success rate can only be determined from a real-system run, we model it at compile time as the {\em program reliablity}. We define the reliability of a program as the product of the reliability of all gates in the program. Although this is not a perfect model for the success rate, it serves as a useful measure of overall correctness \cite{supremacy, linke}. For a given mapping, the solver determines the reliability of each two-qubit and readout operation and computes an overall reliability score. The solver maximizes the reliability score over all mappings by tracking and adapting to the error rates, coherence limits, and qubit movement based on program qubit locations. 

In practice, we use the Z3 SMT solver to express and solve this optimization. Since the reliability objective is a non-linear product, we linearize the objective by optimizing for the additive logarithms of the reliability scores of each gate. We term this algorithm as \rSMTstar. The output of the SMT solver is used to create machine executable code in the vendor-specified assembly language.

\subsection{Real-System Evaluation}
We present real-system evaluation on IBMQ16.
Our evaluation uses 12 common QC benchmarks, compiled using \rSMTstar and \tSMTstar\ which are variants of our compiler and IBM's Qiskit compiler (version 0.5.7) \cite{Qiskit} which is the default for this system. \rSMTstar\ optimizes the reliability of the program using hardware noise data. \tSMTstar\ optimizes the execution time of the program considering real-system gate durations and coherence times, but not operational error rates. IBM Qiskit is also noise-unaware and uses randomized algorithms for SWAP optimization. For each benchmark and compiler, we measured the success rate on IBMQ16 system using 8192 trials per program. Success rate of 1 indicates a perfect noise-free execution. 

Figure \ref{fig:err_full} shows the success rate for the three compilers on all the benchmarks. \rSMTstar\ has higher success rate than both baselines on all benchmarks, demonstrating the effectiveness of noise-adaptive compilation. Across benchmarks \rSMTstar\ obtains geomean $2.9$x improvement over Qiskit, with up to $18$x gain. Figure \ref{fig:bv_all_mappings} shows the mapping used by Qiskit,  \tSMTstar and \rSMTstar for BV4. Qiskit places qubits in a lexicographic order without considering CNOT and readout errors and incurs extra swap operations. Similarly, \tSMTstar is also unaware of noise variations across the device, resulting in mappings which use unreliable hardware. \rSMTstar outperforms these baselines because it maximizes the likelihood of reliable execution by leveraging microarchitectural noise characteristics during compilation.

Full results of our evaluation on 7 QC systems from IBM, Rigetti and UMD can be found in \cite{pmasplos, triq}.

\subsection{Discussion}
Our work represents one of the first efforts to exploit hardware noise characteristics during compilation. 
We developed optimal and heuristic techniques for noise adaptivity and performed comprehensive evaluations on several real QC systems \cite{triq}. We also developed techniques to mitigate crosstalk, another major source of errors in QC systems, using compiler techniques that schedule programs using crosstalk characterization data from the hardware \cite{crosstalk_asplos}. In addition, our techniques are already being used in industry toolflows \cite{qiskit_na, rigetti_na}. Recognizing the importance of efficient compilation, other research groups have also recently developed mapping and routing heuristics \cite{1902.09102, 1712.04722} and techniques to handle noise \cite{1805.10224, Tannu:2019:EDM:3352460.3358257}. 

Our noise-adaptivity optimizations offer large gains in success rate. 
These gains mean the difference between executions which yield correct and usable results and executions where the results are dominated by noise. These improvements are also multiplicative against benefits obtained elsewhere in the stack and will be instrumental in closing the gap between near-term QC algorithms and hardware.
Our work also indicates that it is important to accurately characterize hardware and expose characterization data to software instead of hiding it behind a device-independent ISA layer. Additionally our work also proposes that QC programs should be compiled once-per-execution using the latest hardware characterization data to obtain the best executions. 

Going beyond noise characteristics, we also studied the importance of exposing other microarchitectural information to software. We found that when the compiler has access to the native gates available in the hardware (micro operations used to implement ISA-level gates), it can further optimize programs and improve success rates. Overall, our work indicates that QC machines are not yet ready for technology independent abstractions that shield the software from hardware. Bridging the information gap between software and hardware by breaking abstraction barriers will be increasingly important on the path towards practically useful NISQ devices.
\section{Breaking the Qubit Abstraction via the Third Energy Level}
\label{sec:qutrit}

While quantum computation is typically expressed with the two-level binary abstraction of qubits, the underlying physics of quantum systems are \textbf{not intrinsically binary}. Whereas classical computers operate in binary states at the physical level (e.g. clipping above and below a threshold voltage), quantum computers have natural access to an infinite spectrum of discrete energy levels. In fact, hardware must actively suppress higher level states in order to realize an \textit{engineered} two-level qubit approximation. In this sense, using three-level \textit{qutrits} (quantum trits) is simply a choice of including an additional discrete energy level within the computational space. Thus, it is appealing to explore what gains can be realized by breaking the binary qubit abstraction.

In prior work on qutrits (or more generally, d-level \textit{qudits}), researchers identified only constant factor gains from extending beyond qubits. In general, this prior work \cite{Pavlidis} has emphasized the information compression advantages of qutrits. For example, $N$ qubits can be expressed as $\frac{N}{\log_2(3)}$ qutrits, which leads to $\log_2(3)\approx 1.6$-constant factor improvements in runtimes.

Recently however, our research group demonstrated a novel qutrit approach that leads to exponentially faster runtimes (i.e. shorter in circuit \textit{depth}) than qubit-only approaches \cite{gokhale2019asymptotic, gokhale2020extending}. The key idea underlying the approach is to use the third state of a qutrit as temporary storage. Although qutrits incur higher per-operation error rates than qubits, this is compensated by dramatic reductions in runtimes and quantum gate counts. Moreover, our approach only applies qutrit operations in an intermediary stage: the input and output are still qubits, which is important for initialization and measurement on practical quantum devices \cite{HesingA, HesingB}.

The net result of our work is to extend the frontier of what quantum computers can compute. In particular, the frontier is defined by the zone in which every machine qubit is a data qubit, for example a 100-qubit algorithm running on a 100-qubit machine. In this frontier zone, we do not have space for non-data workspace qubits known as ancilla. The lack of ancilla in the frontier zone is a costly constraint that generally leads to inefficient circuits. For this reason, typical circuits instead operate below the frontier zone, with many machine qubits used as ancilla. Our work demonstrates that ancilla can be substituted with qutrits, enabling us to operate efficiently within the ancilla-free frontier zone.

	\begin{figure}[h]
		\centering
		\includegraphics[width=0.25\textwidth]{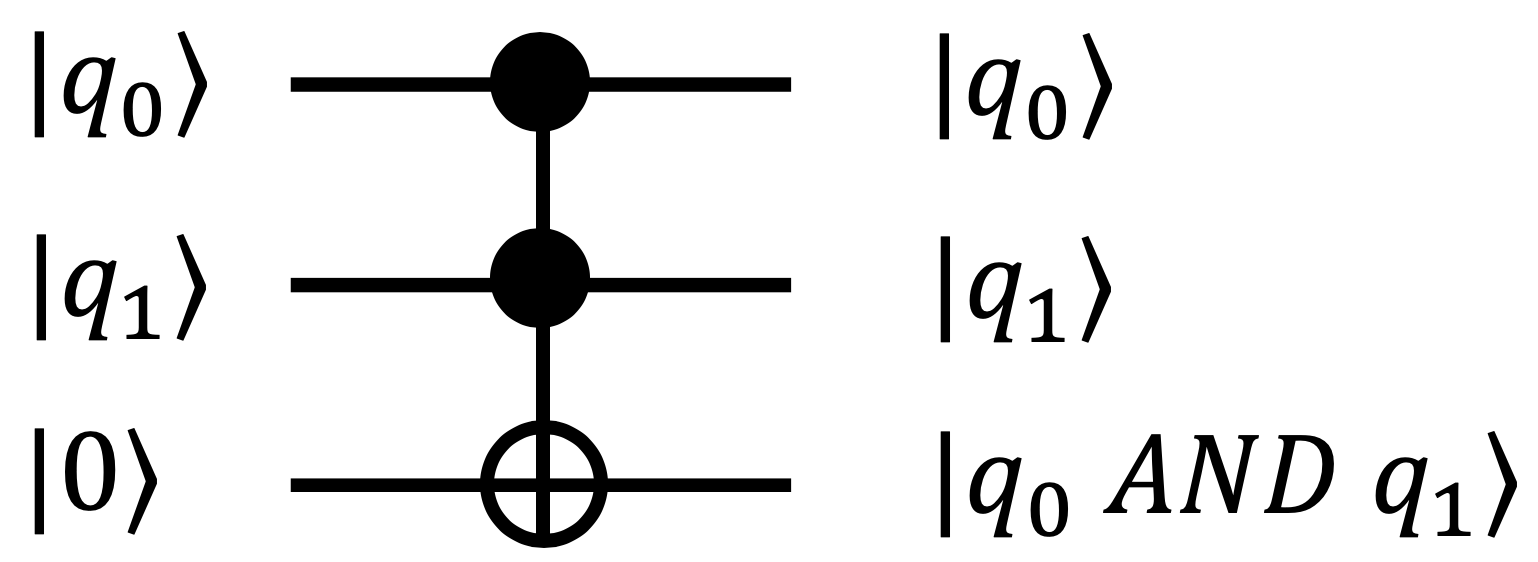}
		\caption{Reversible AND circuit using a single ancilla bit. The inputs are on the left, and time flows rightward to the outputs. This AND gate is implemented using a Toffoli (CCNOT) gate with inputs $q_0$, $q_1$ and a single ancilla initialized to 0. At the end of the circuit, $q_0$ and $q_1$ are preserved, and the ancilla bit is set to 1 if and only if both other inputs are 1.}
		\label{fig:reversible_AND}
	\end{figure}

\subsection{Qutrit-Assisted AND Gate}
We develop the intuition for how qutrits can be useful by considering the example of constructing an AND gate. In the framework of quantum computing, which requires \textit{reversibility}, AND is not permitted directly. For example, consider the output of 0 from an AND gate with two inputs. With only this information about the output, the value of the inputs cannot be uniquely determined (00, 01, and 10 all yield an AND output of 0). However, these operations can be made reversible by the addition of an extra, temporary workspace bit initialized to 0. Using a single additional such \textit{ancilla}, the AND operation can be computed reversibly as in Figure~\ref{fig:reversible_AND}. While this approach works, it is expensive---in order to decompose the \textit{Toffoli} gate in Figure~\ref{fig:reversible_AND} into hardware-implementable one- and two- input gates, it is decomposed into at least six CNOT (controlled-NOT) gates.

However, if we break the qubit abstraction and allow occupation of a higher \textit{qutrit} energy level, the cost of the Toffoli AND operation is greatly diminished. Before proceeding, we review the basics of qutrits, which have three computational basis states: $\ket{0}$, $\ket{1}$, and $\ket{2}$. A qutrit state $\ket{\psi}$ may be represented analogously to a qubit as $\ket{\psi} = \alpha\ket{0} + \beta\ket{1} + \gamma\ket{2}$, where $\norm{\alpha}^2 + \norm{\beta}^2 + \norm{\gamma}^2 = 1$. Qutrits are manipulated in a similar manner to qubits; however, there are additional gates which may be performed on qutrits. We focus on the $X_{+1}$ and $X_{-1}$ operations, which are addition and subtraction gates, modulo 3. For example $X_{+1}$ elevates $\ket{0}$ to $\ket{1}$ and elevates $\ket{1}$ to $\ket{2}$, while wrapping $\ket{2}$ to $\ket{0}$.

Just as single-qubit gates have qutrit analogs, the same holds for two-qutrit gates. For example, consider the CNOT operation, where an X gate is performed conditioned on the control being in the $\ket{1}$ state. For qutrits, an $X_{+1}$ or $X_{-1}$ gate may be performed, conditioned on the control being in any of the three possible basis states. Just as qubit gates are extended to take multiple controls, qutrit gates are extended similarly.

\begin{figure}[h]
\centering
\includegraphics[width=0.3\textwidth]{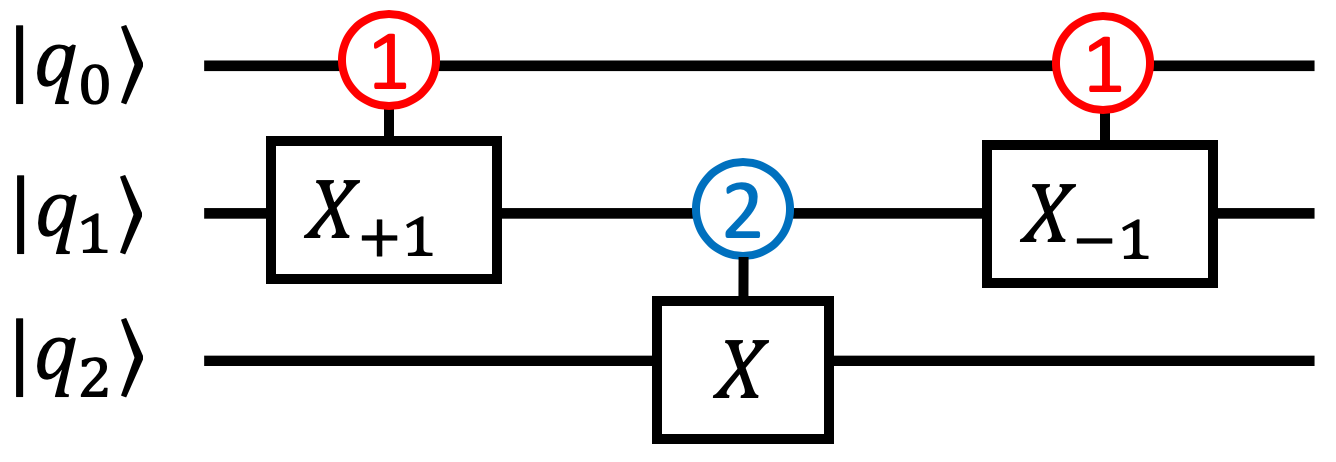}
\caption{A Toffoli decomposition via qutrits. Each input and output is a qubit. The red controls activate on $\ket{1}$ and the blue controls activate on $\ket{2}$. The first gate temporarily elevates $q_1$ to $\ket{2}$ if both $q_0$ and $q_1$ were $\ket{1}$. We then perform the X operation only if $q_1$ is $\ket{2}$. The final gate restores $q_0$ and $q_1$ to their original state.}
\label{fig:toffoli_decomposition}
\end{figure}

In Figure \ref{fig:toffoli_decomposition}, a Toffoli decomposition using qutrits is given. A similar construction for the Toffoli gate is known from past work \cite{Lanyon, Ralph}. The goal is to perform an X operation on the last (target) input qubit $q_2$ if and only if the two control qubits, $q_0$ and $q_1$, are both $\ket{1}$. First a $\ket{1}$-controlled $X_{+1}$ is performed on $q_0$ and $q_1$. This elevates $q_1$ to $\ket{2}$ iff $q_0$ and $q_1$ were both $\ket{1}$. Then a $\ket{2}$-controlled $X$ gate is applied to $q_2$. Therefore, $X$ is performed only when both $q_0$ and $q_1$ were $\ket{1}$, as desired. The controls are restored to their original states by a $\ket{1}$-controlled $X_{-1}$ gate, which undoes the effect of the first gate. The key intuition in this decomposition is that the qutrit $\ket{2}$ state can be used instead of ancilla to store temporary information.

\subsection{Generalized Toffoli Gate}
\label{subsec:generalized_toffoli_construction}

\begin{figure}[h]
  \includegraphics[width=0.48\textwidth]{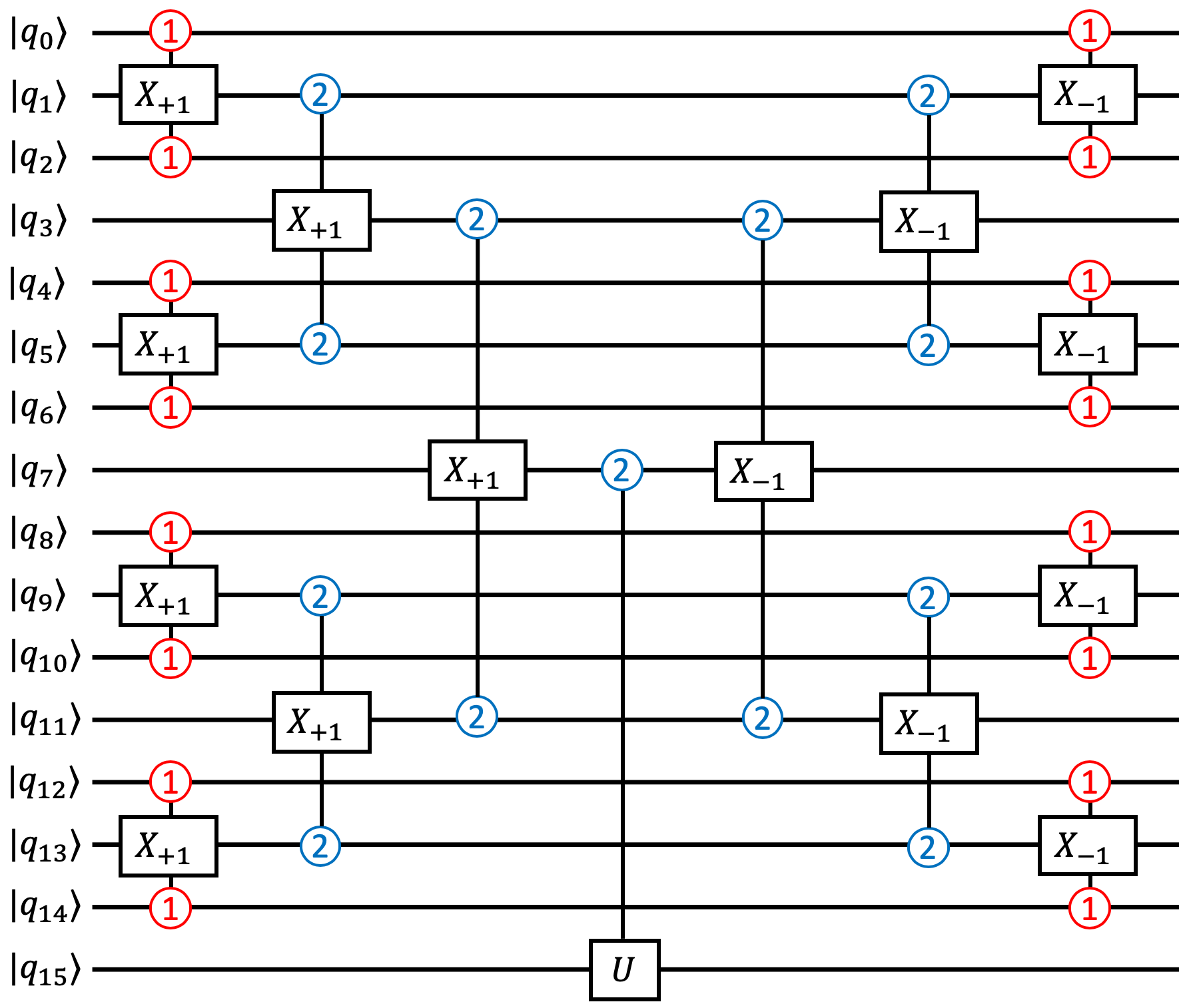}
  \caption{Our circuit decomposition for the Generalized Toffoli gate is shown for 15 controls and 1 target. The inputs and outputs are both qubits, but we allow occupation of the $\ket{2}$ qutrit state in between. The circuit has a tree structure and maintains the property that the root of each subtree can only be elevated to $\ket{2}$ if all of its control leaves were $\ket{1}$. Thus, the $U$ gate is only executed if all controls are $\ket{1}$. The right half of the circuit performs uncomputation to restore the controls to their original state. This construction applies more generally to any multiply-controlled $U$ gate. Note that the three-input gates are decomposed into 6 two-input and 7 single-input gates in our actual simulation, as based on the decomposition in \cite{Di}.}
  \label{fig:btb_cnu}
\end{figure}
The intuition of our technique extends to more complicated gates. In particular, we consider the Generalized Toffoli gate, a ubiquitous quantum operation which extends the Toffoli gate to have any number of control inputs. The target input is flipped if and only if all of the controls are activated. Our qutrit-based circuit decomposition for the Generalized Toffoli gate is presented in Figure~\ref{fig:btb_cnu}. The decomposition is expressed in terms of three-qutrit gates (two controls, one target) instead of single- and two- qutrit gates, because the circuit can be understood purely classically at this granularity. In actual implementation and in our simulation, we used a decomposition \cite{Di} that requires 6 two-qutrit and 7 single-qutrit physically implementable quantum gates.

Our circuit decomposition is most intuitively understood by treating the left half of the the circuit as a tree. The desired property is that the root of the tree, $q_7$, is $\ket{2}$ if and only if each of the 15 controls was originally in the $\ket{1}$ state. To verify this property, we observe the root $q_7$ can only become $\ket{2}$ iff $q_7$ was originally $\ket{1}$ and $q_3$ and $q_{11}$ were both previously $\ket{2}$. At the next level of the tree, we see $q_3$ could have only been $\ket{2}$ if $q_3$ was originally $\ket{1}$ and both $q_1$ and $q_5$ were previously $\ket{2}$, and similarly for the other triplets. At the bottom level of the tree, the triplets are controlled on the $\ket{1}$ state, which are only activated when the even-index controls are all $\ket{1}$. Thus, if any of the controls were not $\ket{1}$, the $\ket{2}$ states would fail to propagate to the root of the tree. The right half of the circuit performs \textit{uncomputation} to restore the controls to their original state.

After each subsequent level of the tree structure, the number of qubits under consideration is reduced by a factor of $\sim2$. Thus, the circuit depth is logarithmic in $N$, which is \textbf{exponentially smaller than ancilla-free qubit-only circuits}. Moreover, each qutrit is operated on by a constant number of gates, so the total number of gates is linear in $N$.

We verified our circuits, both formally and via simulation. Our verification scripts are available on our GitHub \cite{QutritsGithub}.

\subsection{Simulation Results}

\begin{table}[h] \renewcommand{\arraystretch}{1.5}
\centering
\begin{tabular}{l|ccc}
           & QUBIT & QUBIT+ANCILLA & QUTRIT \\ \hline
Depth      & $\sim 633N$     & $\sim 76N$             & $\sim 38 \log_2(N)$      \\
Gate Count & $\sim 397N$     & $\sim 48N$             & $\sim 6N$     
\end{tabular}
\caption{Scaling of circuit depths and two-qudit gate counts for all three benchmarked circuit constructions for the $N$-controlled Generalized Toffoli.} \label{tab:costs}
\end{table}

\begin{figure*}[ht]
    \centering
        \begin{tikzpicture}[baseline]
\pgfplotsset{every axis/.append style={thick}}
\pgfplotsset{every tick label/.append style={font=\small}}
\pgfplotsset{every axis label/.append style={font=\small}}

\newcommand{\extraBarHeight}{1}

\begin{axis}[
    title=Fidelity for Superconducting Models,
    symbolic x coords={left,SC,a,SC+T1,b,SC+GATES,c,SC+T1+GATES,right},
    width=.66\textwidth,
    height=150pt,
    ybar=5pt,
    bar width=15pt,
    xmin=left, xmax=right,
    ymin=0, ymax=120,
    ytick={0, 25, 50, 75, 100},
    ytick distance=25,
    xtick=data,
    ,
    legend style={draw=none, fill=none, at={(0.76,1.03)},anchor=north,font=\small},
    legend columns=-1,
    legend image code/.code={\draw[#1, draw=none] (0em,-0.2em) rectangle (0.6em,0.4em);},
    axis line style={draw=black!20!white},
    axis on top,
    y axis line style={draw=none},
    axis x line*=bottom,
    tick style={draw=none},
    yticklabel={\pgfmathparse{\tick*1}\pgfmathprintnumber{\pgfmathresult}\%},
    clip=false,
    enlarge y limits=0,
    ,
    nodes near coords always on top/.style={
        every node near coord/.append style={
            anchor=south,
            rotate=0,
            font=\small,
            inner sep=0.2em,
        },
    },
    nodes near coords={
        \StrPosition{\pgfplotspointmeta}{Y}[\Result]%
        \StrGobbleLeft{\pgfplotspointmeta}{\Result}[\Result]%
        \StrGobbleRight{\Result}{1}[\Result]%
        \pgfmathparse{\Result<10}%
        \ifnum\pgfmathresult=1
            \pgfmathparse{\Result-\extraBarHeight}%
        \else
            \pgfmathparse{\Result}%
        \fi%
        \pgfmathprintnumber[fixed,precision=2]{\pgfmathresult}\%
    },
    nodes near coords always on top,
]

\addplot[style={color=transparent, draw=none, fill=\colorQubitLight, mark=none}] coordinates {
    (SC, 1.01)  
    (SC+T1, 1.56)  
    (SC+GATES, 1.01)  
    (SC+T1+GATES, 26.1)  
};
\addlegendentry{QUBIT\:\:\:\:};

\addplot[style={draw=none, pattern color=\colorQubitBB, pattern=north east lines, mark=none}] coordinates {
    (SC, 18.5)  
    (SC+T1, 52.3)  
    (SC+GATES, 30.2)  
    (SC+T1+GATES, 84.1)  
};
\addlegendentry{QUBIT+ANCILLA\:\:\:\:};

\addplot[style={draw=none, fill=\colorQutritLight, mark=none}] coordinates {
    (SC, 56.8)  
    (SC+T1, 65.9)  
    (SC+GATES, 83.1)  
    (SC+T1+GATES, 94.7)  
};
\addlegendentry{QUTRIT};

\end{axis}

%
%
%
%
\end{tikzpicture}%
%
        \:\:\:\:\:\:\:\:
        \begin{tikzpicture}[baseline]
\pgfplotsset{every axis/.append style={thick}}
\pgfplotsset{every y tick label/.append style={font=\small}}
\pgfplotsset{every x tick label/.append style={font=\small}}
\pgfplotsset{every axis label/.append style={font=\small}}

\newcommand{\beginCustomBarAxisFirst}[1]{
\begin{axis}[
    title=Fidelity for Trapped Ion Models,
    symbolic x coords={#1},
    width=0.33\textwidth,
    height=150pt,
    ybar=5pt,
    bar width=15pt,
    xmin=left, xmax=right,
    ymin=0, ymax=120,
    ytick={0, 25, 50, 75, 100},
    ytick distance=25,
    xtick=data,
    enlarge x limits=0,
    ,
    legend style={draw=none, fill=none, at={(0.5,1.03)},anchor=north,font=\small},
    legend columns=2,
    legend image code/.code={\draw[##1, draw=none] (0em,-0.2em) rectangle (0.6em,0.4em);},
    axis line style={draw=black!20!white},
    axis on top,
    y axis line style={draw=none},
    axis x line*=bottom,
    tick style={draw=none},
    yticklabel={~},
    clip=false,
    enlarge y limits=0,
    ,
    nodes near coords always on top/.style={
        every node near coord/.append style={
            anchor=south,
            rotate=0,
            font=\small,
            inner sep=0.2em,
        },
    },
    nodes near coords={
        \pgfmathprintnumber[fixed,fixed zerofill,precision=1]{\pgfplotspointmeta}\%
    },
    nodes near coords always on top,
]
}

\newcommand{\beginCustomBarAxis}[1]{
\begin{axis}[
    title=,
    symbolic x coords={#1},
    width=0.33\textwidth,
    height=150pt,
    ybar=5pt,
    bar width=15pt,
    xmin=left, xmax=right,
    ymin=0, ymax=120,
    ytick={0, 25, 50, 75, 100},
    ytick distance=25,
    xtick=data,
    enlarge x limits=0,
    x tick label style={rotate=90, anchor=west, yshift=-0.1em, xshift=0.4em, color=black},
    ,
    legend style={draw=none, fill=none, at={(0.5,1.03)},anchor=north,font=\small},
    legend columns=2,
    legend image code/.code={\draw[##1, draw=none] (0em,-0.2em) rectangle (0.6em,0.4em);},
    axis line style={draw=none},
    axis on top,
    tick style={draw=none},
    yticklabel={~},
    clip=false,
    enlarge y limits=0,
    ,
    nodes near coords always on top/.style={
        every node near coord/.append style={
            anchor=south,
            rotate=0,
            font=\small,
            inner sep=0.2em,
        },
    },
    nodes near coords={
        \pgfmathprintnumber[fixed,fixed zerofill,precision=1]{\pgfplotspointmeta}\%
    },
    nodes near coords always on top,
]
}

\beginCustomBarAxisFirst{left,q,TI\textunderscore QUBIT,a,b,x,c,d,y,right}
\addplot[style={draw=none, fill=\colorQubitLight, mark=none}] coordinates {
    (TI\textunderscore QUBIT, 44.66)
};

\addplot[style={draw=none, pattern color=\colorQubitBB, pattern=north east lines, mark=none}] coordinates {
    (TI\textunderscore QUBIT, 89.85)
};
\end{axis}

\beginCustomBarAxis{left,q,v,u,c,d,BARE\textunderscore QUTRIT,e,f,y,right}
\addplot[style={draw=none, fill=\colorQutritLight, mark=none}] coordinates {
    (BARE\textunderscore QUTRIT, 94.92)
};
\end{axis}

\beginCustomBarAxis{left,q,u,c,d,x,e,f,DRESSED\textunderscore QUTRIT,right}
\addplot[style={draw=none, fill=\colorQutritLight, mark=none}] coordinates {
    (DRESSED\textunderscore QUTRIT, 96.08)
};
\end{axis}

\end{tikzpicture}
    \caption{Circuit simulation results for all possible pairs of circuit constructions and noise models. Each bar represents 1000+ trials, so the error bars are all $2\sigma < 0.1\%$. Our QUTRIT construction significantly outperforms the QUBIT construction. The QUBIT+ANCILLA bars are drawn with dashed lines to emphasize that it has access to an extra ancilla bit, unlike our construction. Figure reprinted with permission from \cite{gokhale2019asymptotic}}
    \label{fig:simulation_results}
\end{figure*}
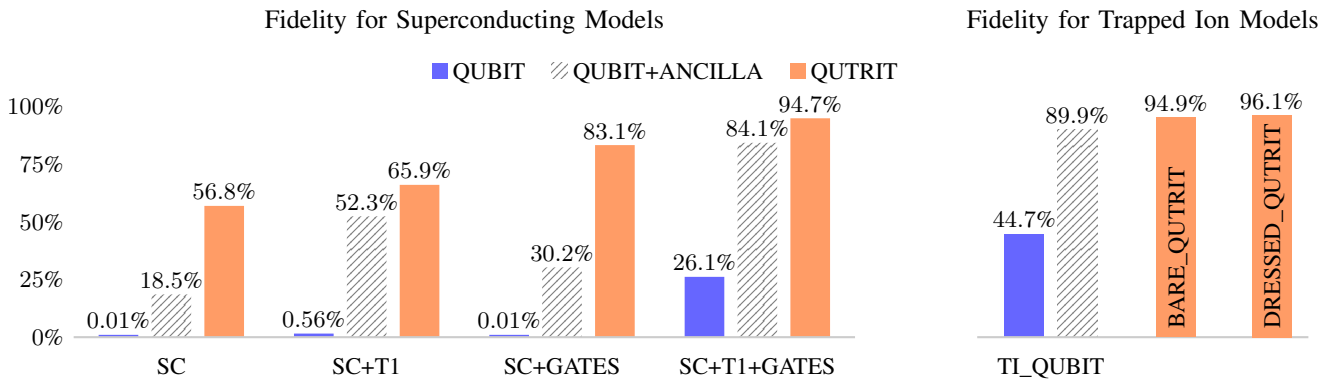

Table~\ref{tab:costs} shows the scaling of circuit depths and two-qudit gate counts for all three benchmarked circuits. The QUBIT-based circuit constructions from past work are linear in depth and have a high linearity constant. Augmenting with a single borrowed ancilla (QUBIT+ANCILLA) reduces the circuit depth by a factor of 8. However, both circuit constructions are significantly outperformed by our QUTRIT construction, which scales logarithmically in $N$ and has a relatively small leading coefficient. While there is not an asymptotic scaling advantage for two-qudit gate count, the linearity constant for our QUTRIT circuit is 70x smaller than for the equivalent ancilla-free QUBIT circuit.

We ran simulations under realistic superconducting and trapped ion device noise. The simulations were run in parallel on over 100 n1-standard-4 Google Cloud instances. These simulations represent over 20,000 CPU hours, which was sufficient to estimate mean fidelity to an error of $2 \sigma < 0.1\%$ for each circuit-noise model pair.

The full results of our circuit simulations are shown in Figure~\ref{fig:simulation_results}. All simulations are for the 14-input (13 controls, 1 target) Generalized Toffoli gate. We simulated each of the three circuit benchmarks against each of our noise models (when applicable), yielding the 16 bars in the figure. Notice that our qutrit circuit consistently outperforms qubit circuits, with advantages ranging from 2x to 10,000x.

\subsection{Discussion}
The results presented in our work in \cite{gokhale2019asymptotic, gokhale2020extending} are applicable to quantum computing in the near term, on machines that are expected within the next five years. By breaking the qubit abstraction barrier, we extend the frontier of what is computable by quantum hardware right now, without needing to wait for better hardware. As verified by our open-source circuit simulator coupled with realistic noise models, our circuits are more reliable than qubit-only equivalents, suggesting that qutrits offer a promising path towards scaling quantum computers. We propose further investigation into what advantage qutrits or qudits may confer. More broadly, it is critical for quantum architects to bear in mind that standard abstractions in classical computing do not necessarily transfer to quantum computation. Often, this presents unrealized opportunities, as in the case of qutrits.

\begin{figure*}[h]
    \centering
    \includegraphics[height=0.215\textwidth]{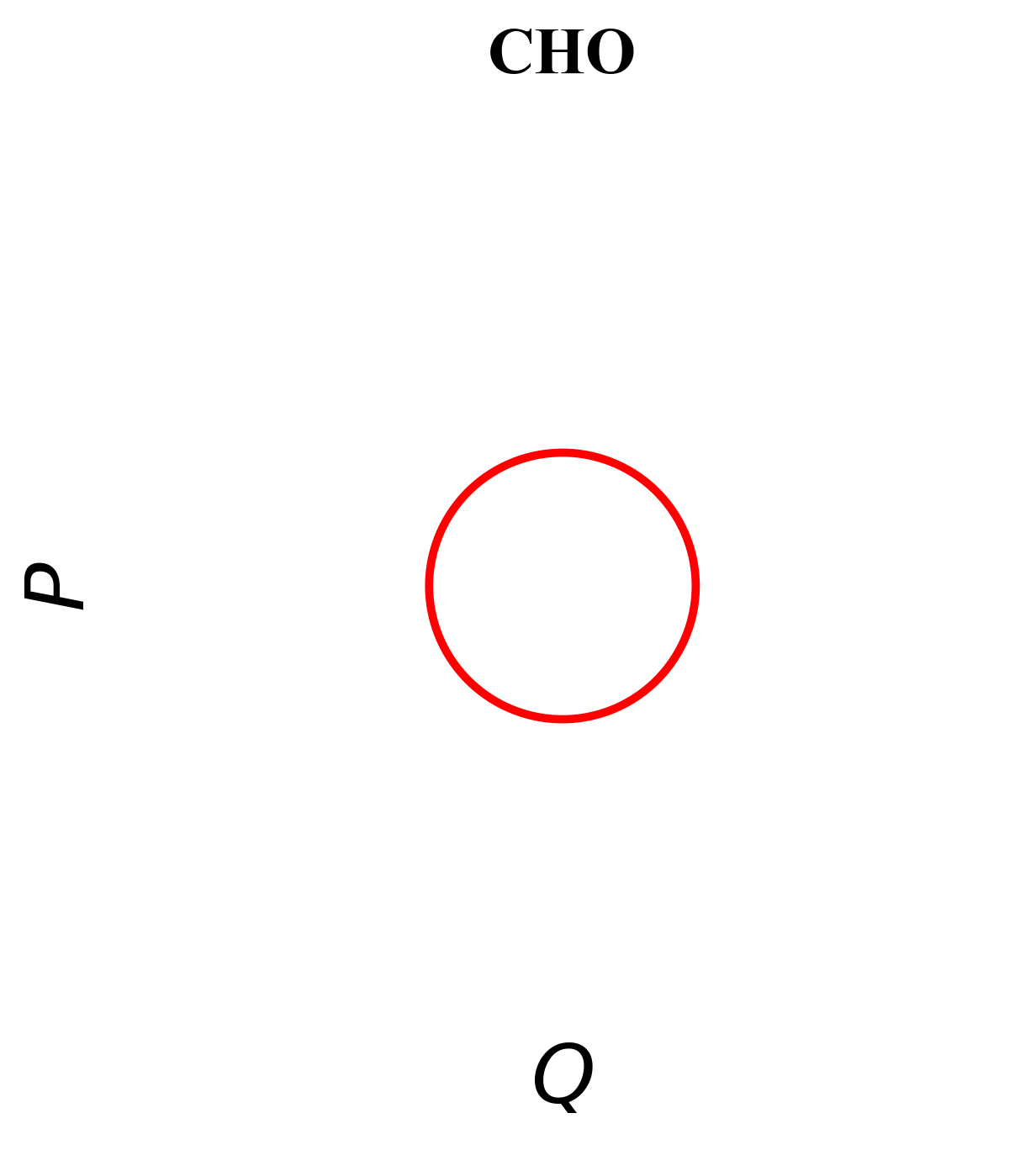}
     \includegraphics[height=0.215\textwidth]{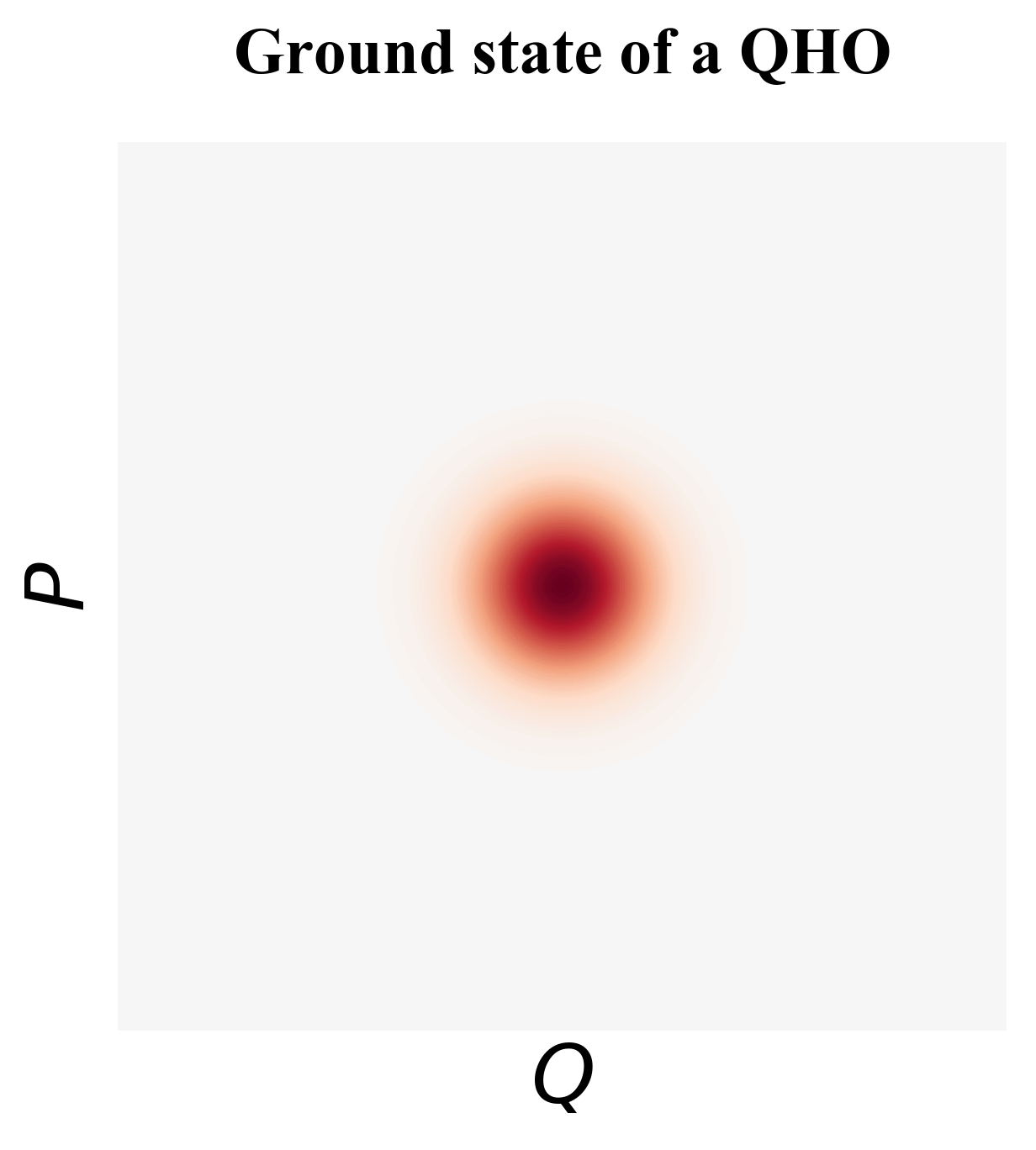}
      \includegraphics[height=0.215\textwidth]{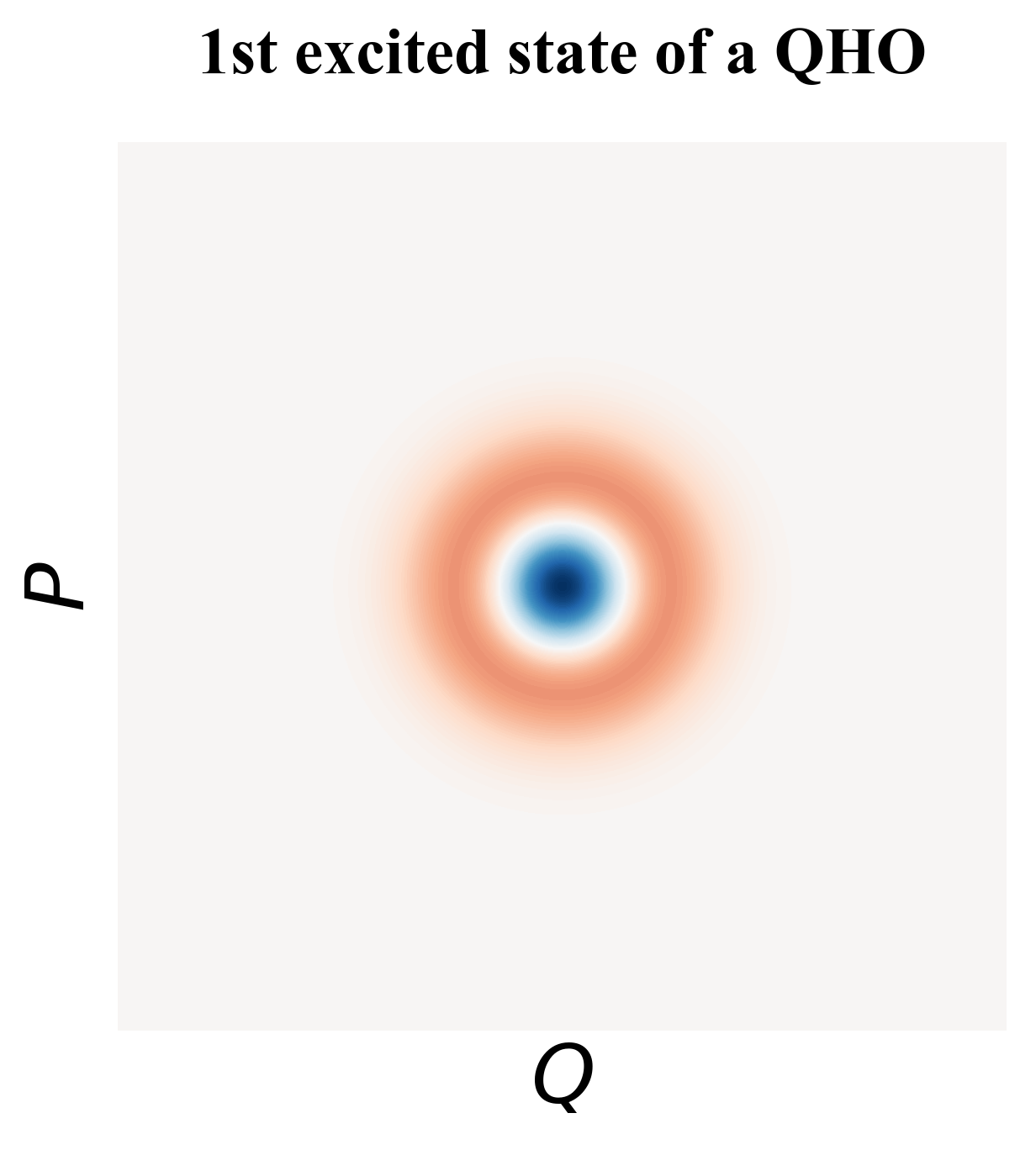}\includegraphics[height=0.215\textwidth]{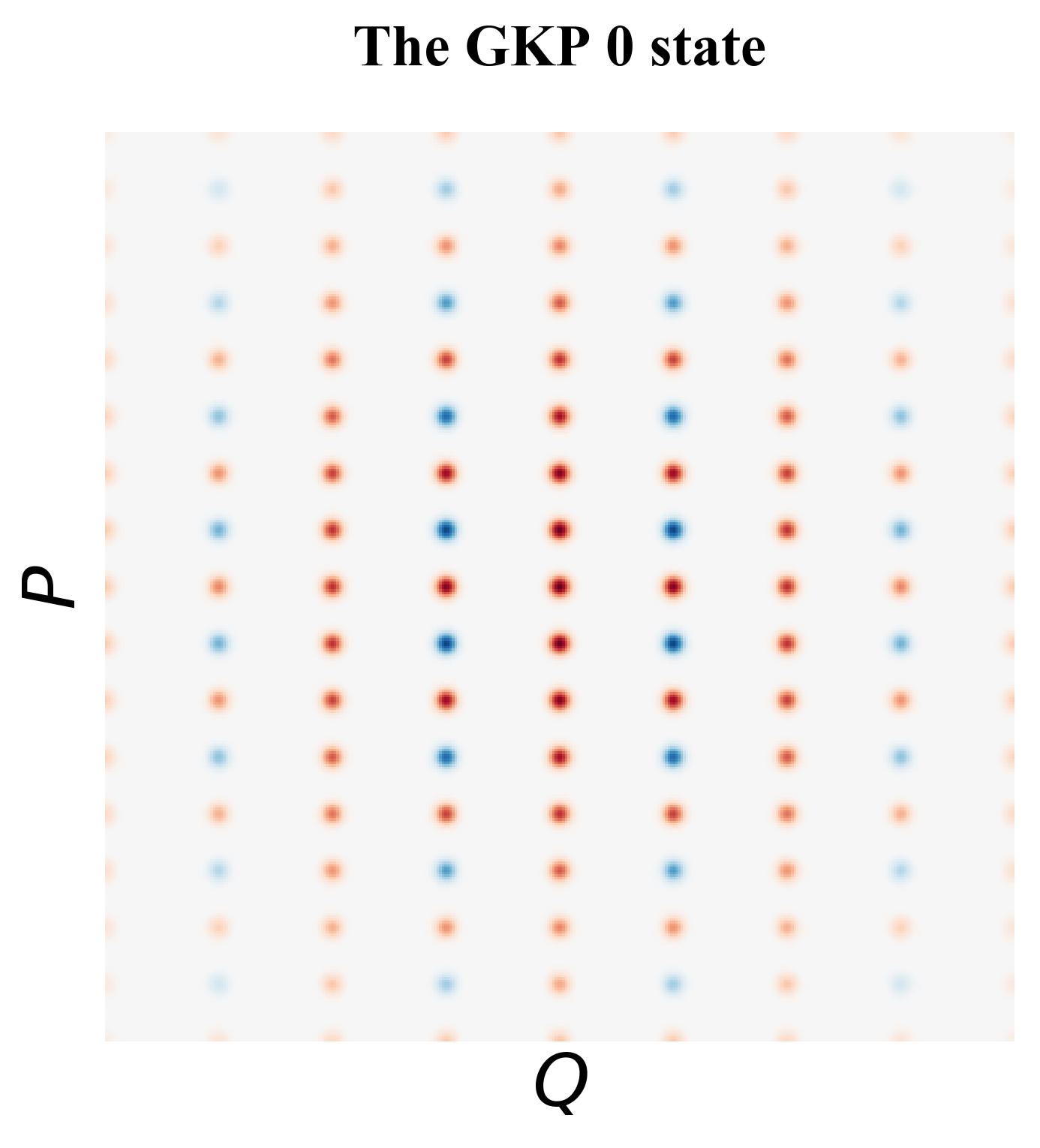}\includegraphics[height=0.215\textwidth]{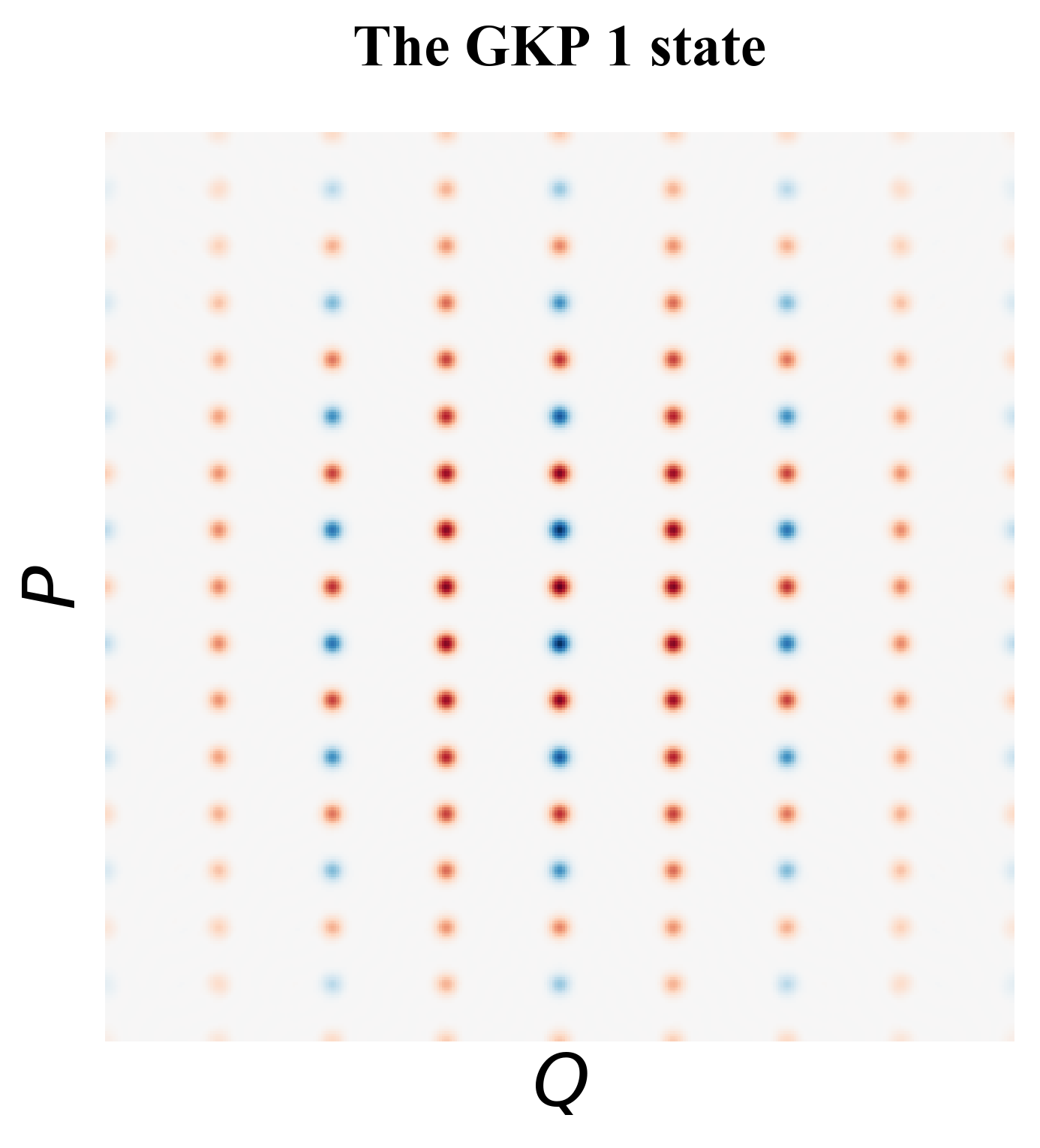}
    \caption{Phase space diagrams for a Classical Harmonic Oscillator (CHO), the ground state and the first excited state of a Quantum Harmonic Oscillator (QHO) and the logic 0 and 1 state of the GKP qubit. For quantum phase space diagrams, the plotted distribution is the Wigner quasi-probability function, where red indicates positive values and blue indicates negative values.}
    \label{fig:wig}
\end{figure*}
\section{Breaking the Qubit Abstraction via the GKP Encoding}
\label{sec:gkp}
Currently, there are many competing physical qubit implementations. For example, the transmon qubits \cite{Koch2007} are encoded in the lowest two energy levels of the charge states in superconducting LC circuits with Josephson junctions; Trapped ion qubits can be encoded in two ground state hyperfine levels \cite{Blinov2005} or a ground state level and an excited level of an ion \cite{Cirac1995}; quantum dot qubits use electron spin triplets \cite{Loss1998}. These QIP platfroms have rather distinct physical characteristics, but they are all exposed to the other layers in the stack as qubits and other implementation details are often hidden. This abstraction is natural for classical computing stack because that the robustness of classical bits decouples the programming logic from physical properties of the transistors except the logical value. In contrast, qubits are fragile so there are more than (superpositions of) the logical values that we want to know about the implementation. For example, in the transmon qubits and trapped ion qubits, logical states can be transferred to higher levels of the physical space by unwanted operations and this can cause leakage errors \cite{Ghosh2013, Wood2018}. It will be useful for other layers in the stack to access this error information and develop methods to mitigate it. In the previous section we discussed the qutrit approach that directly uses the third level for information processing, however, it could be more interesting if we can encode the qubit (qudit) using the whole physical Hilbert space to avoid leakage errors systematically and use the redundant degrees of freedom to reveal information about the noise in the encoding. The encoding proposed by Gottesman, Kitaev and Preskill (GKP) \cite{GKP2001} provides such an example. GKP encoding is free of leakage errors and other errors (in the form of small shifts in phase space) can be identified and corrected by Quantum Non-Demolition (QND) measurements and simple linear optical operations. In realistic implementations of approximate GKP states (\cref{sec:gkp_encoding}), there are leakage errors between logical states, but the transfer probability is estimated to be at the order of $10^{-10}$ with current techonology, thus negligible.

\subsection{The phase space diagram}
We describe the GKP qubits in the phase space. For a comparison, we first discuss the phase space diagram for a classical harmonic oscillator and a superconducting qubit. 

\textbf{Classical Harmonic Oscillators} Examples of Classical Harmonic Oscillators (CHO) include LC circuits, springs and pendulums with small displacement. The voltage/displacement (denoted as $p$) and the current/momentum (denoted as $q$) value completely characterize the dynamics of CHO systems. The phase space diagram plots $p$ vs $q$, which for CHOs are circles (up to normalization) with the radius representing the system energy. The energy of CHOs can be any non-negative real value.

\textbf{Quantum Harmonic Oscillators} 
The Quantum Harmonic Oscillator is the quantized verision of the CHO and is the physical model for superconducting LC circuits and superconducting cavity modes. One of the values get quantized for QHOs is the system energy, which can only take equally-spaced non-zero discrete values. The lowest allowed energy is not 0 but $
\frac{1}{2}$ (up to normalization). We call the quantum state with the lowest energy the ground state. For a motion with a certain energy, the phase space diagram is not a circle anymore but a quasi-distribution that can be described by the Wigner function. We say the distribution is a ``quasi" distribution because the probability can be negative. The phase space diagram for the ground state and first excited state is plot in \cref{fig:wig}.

\begin{figure}[h]
    \centering
    \includegraphics[height=0.12\textwidth]{figs/lc.png}
    $\ \ \ \ \ \ $ \includegraphics[height=0.1\textwidth]{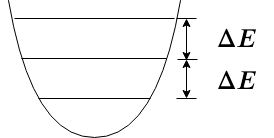}
    \caption{Left: an LC circuit. In superconducting LC circuits, normal current becomes superconducting current. Right: the energy potential of a Harmonic Oscillator. In QHOs like the superconducting LC circuits, the system energy becomes equally spaced discrete values. The plotted two levels are the ground state and the first excited state}
    \label{fig:qho}
\end{figure}

\begin{figure}[h]
    \centering
    \includegraphics[height=0.12\textwidth]{figs/jj.png}
    $\ \ \ \ \ \ $ \includegraphics[height=0.1\textwidth]{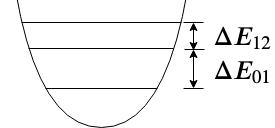}
    \caption{Left: an LC circuit. In superconducting LC circuits, normal current becomes superconducting current. Right: the energy potential of a Harmonic Oscillator. In QHOs like the superconducting LC circuits, the system energy becomes equally spaced discrete values. The plotted two levels are the ground state and the first excited state.}
    \label{fig:jj}
\end{figure}

\textbf{Superconducting Charge Qubits} The QHO does not allow us selectively address the energy levels, thus leakage errors will occur if we use the lowest two levels as the qubit logic space. For example, a control signal that provides the energy difference $\Delta E$ enables the transition $\ket{0}\rightarrow\ket{1}$, but will also make the transition $\ket{1}\rightarrow\ket{2}$ which brings the state out of the logic space.  To avoid this problem, the Cooper Pair Box (CPB) design of a superconducting charge qubit replaces the inductor (see \cref{fig:jj}) with a Josephson junction, making the circuit an anharmonic oscillator, in which the energy levels are not equally spaced anymore. The Wigner function for CPB eigenstates are visually similar to those of QHO and only differ from them to the first order of the anharmonicity, thus we do not plot them in \cref{fig:wig} separately.

\subsection{The Heisenberg uncertainty principle} 
\begin{figure}[h]
    \centering
    \includegraphics[height=0.22\textwidth]{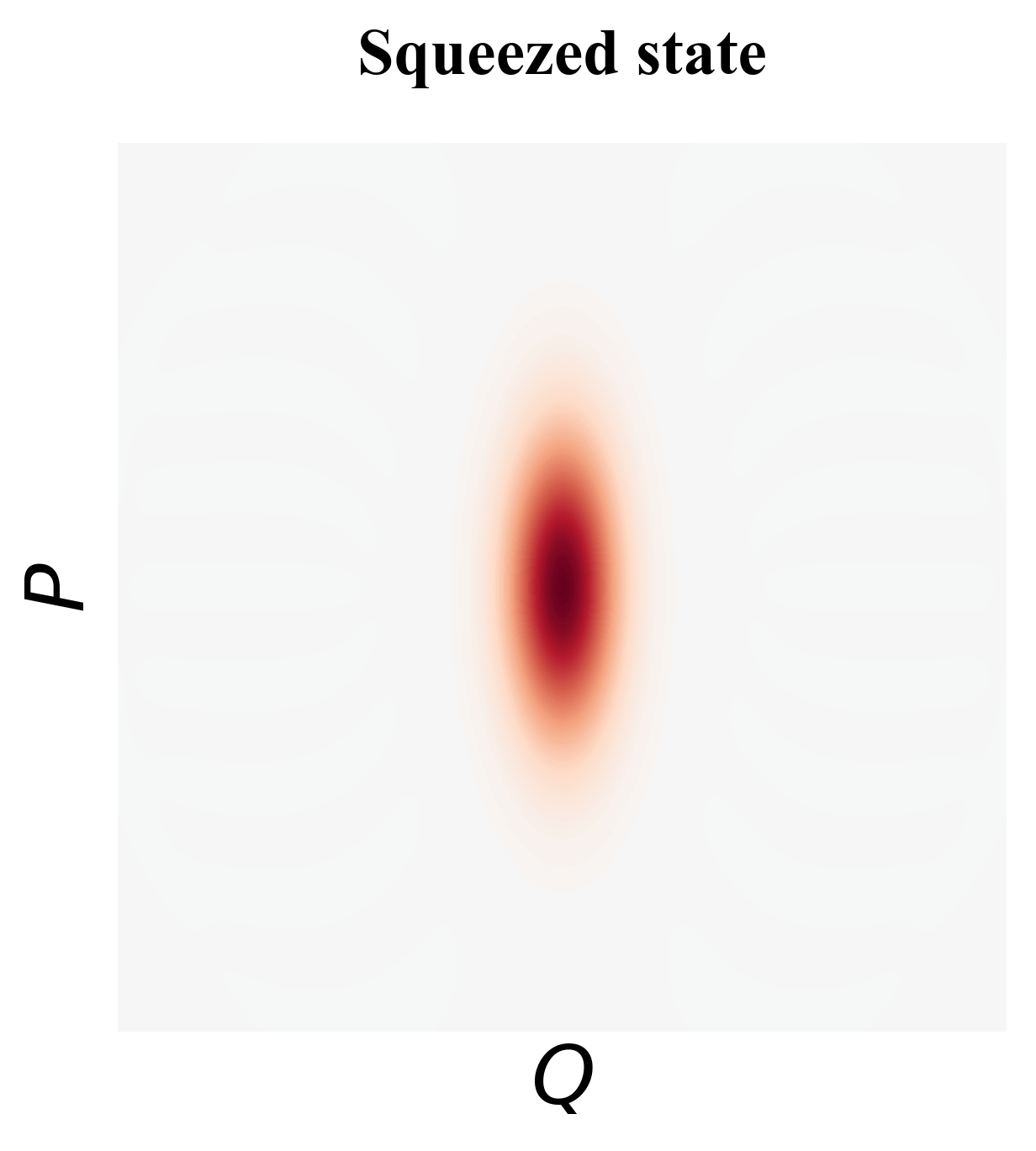}
    \caption{A squeezed vacuum state}
    \label{fig:sqz}
\end{figure}

We hope that with utilizing the whole physical states (higher energy levels), we can use the redundant space to encode and extract error information. However, the Heisenberg uncertainty principle sets the fundamental limit on what error information we can extract from the physical states --- the more we know about the $q$ variable, the less we know about the $p$ variable. For example, we can ``squeeze" the ground state of the QHO (also known as the vacuum state) in the $p$ direction, however, the distribution in  the $q$ direction spreads, as shown in \cref{fig:sqz}. Usually, we have to know both the $p$ and $q$ value to characterize the error information unless we know the error is biased. Thus, it's a great challenge to design encodings in the phase space to reveal error information.

\subsection{The GKP encoding}
\label{sec:gkp_encoding}
The GKP states are also called the grid states because each of them is a rectangular lattice in the phase space (see \cref{fig:wig}). There are also other types of lattice in the GKP family, for example, the hexagonal GKP \cite{GKP2001}. Intuitively, the GKP encoding ``breaks" the Heisenberg uncertainty principle---we do not know what are the measured $p$ and $q$ values of the state (thus expected values of $p$ and $q$ remain uncertain), but we do know that they must be integer multiples of the spacing of the grid. Thus, we have access to the error information in both directions and if we measure values that are not multiples of the spacing of the grid, we know there must be errors. Formally, the ideal GKP logical states are given by, 
\begin{align}
\ket{\overline{0}}_{\text{gkp}} &= \sum_{k= - \infty}^{\infty} S_{p}^{k} \ket{q=0} \nonumber \\
\ket{\overline{1}}_{\text{gkp}} &=\sum_{k= - \infty}^{\infty} S_{p}^{k} \ket{q=\sqrt{\pi}},
\label{eq:ExactGKPstates}
\end{align}

where $S_p = e^{-2i\sqrt{\pi} p}$ is the displacement operator in $q$ space, which shift a wave function in the $q$ direction by 2$\sqrt{\pi}$. These definitions show that for GKP logical 0 and 1, the spacing of the grid in $q$ direction is $2\sqrt{\pi}$ and the spacing in the $p$ is $\sqrt{\pi}$. In $q$ direction, the logical $\ket{0}$ state has peaks at even multiples of $\sqrt{\pi}$ and the logical $\ket{1}$ state has peaks at odd multiples of $\sqrt{\pi}$. For logical $\ket{+}$ and $\ket{-}$, the spacing in $p$ and $q$ grid s are switched.

\begin{figure}[h]
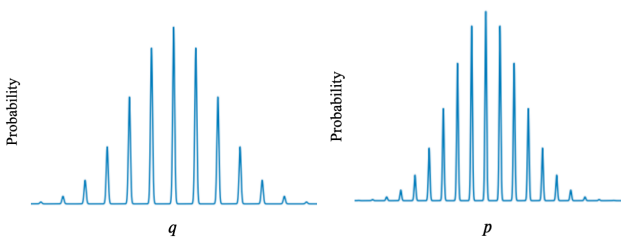

    \centering
    \includegraphics[height=0.16\textwidth]{figs/gkp_q.png}
    \includegraphics[height=0.17\textwidth]{figs/gkp_p.png}
    \caption{Approximate GKP $
\ket{0}$  state  in $q$ and $p$ axis.}
    \label{fig:approx_gkp}
\end{figure}

\textbf{Approximate GKP states} The ideal GKP states require infinite energy thus are not realistic. In the lab, we can prepare approximate GKP states as illustrated in \cref{fig:approx_gkp}, where peaks and the envelope are Gaussian curve.

\textbf{Error correction with GKP qubits} 
GKP qubits are designed to correct shift errors in $q$ and $p$ axis. A simple decoding strategy will be shifting the GKP state back to the closest peak. For example, if we measure a $q$ value to be $2\sqrt{\pi}+\Delta q$, where $\Delta q < \frac{\sqrt{\pi}}{2}$, then we can shift it back to $2\sqrt{\pi}$. With this simple decoding, GKP can correct all shift errors smaller than $\frac{\sqrt{\pi}}{2}$. While there are other proposals for encoding qubits in QHO \cite{catcode, GaussianChannel18, Michael2016} that are designed for realistic errors such as photon loss, it is shown that GKP qubits have the most error correcting ability in the regime of experimental relevance \cite{AlbertBosonicCodePerformance18}. 

In addition, GKP qubits can also provide error correction information when concatenating with Quantum Error Correction Codes (QECC) and yield higher thresholds. For example, when combing the GKP qubits with a surface code, the measured continuous $p$ and $q$ value in the stabilizer measurement can reveal more about the error distribution than traditional qubits \cite{Noh2019, Vuillot2019, Fukui2018}.

Lastly, it has been shown that given a supply of GKP-encoded Pauli eigenstates, universal fault-tolerant quantum computation can be achieved using only Gaussian operations \cite{MagicStateGKP19}. Comparing to qubit error correction codes, the GKP encoding enables much simpler fault-tolerant constructions.

\subsection{Fault-Tolerant Preparation of Approximate GKP States}
 The GKP encoding has  straightforward logical operation and  promising error correcting performance. However, the difficulty of using GKP qubits in QIP platforms lies in its preparation since they live in highly non-classical states with relatively high mean photon number ($i.e.,$ the average energy levels). Thus, realiable preparation of encoded GKP states is an important problem. In \cite{shiNJP2019}, we gave fault-tolerance definitions for GKP preparation in superconducting cavities and designed a protocol that fault-tolerantly prepares the GKP states. We briefly describe the main ideas.
     \subsubsection{Goodness of approximate GKP states}

\begin{figure}[h]
\centering
\includegraphics[width=0.48\textwidth]{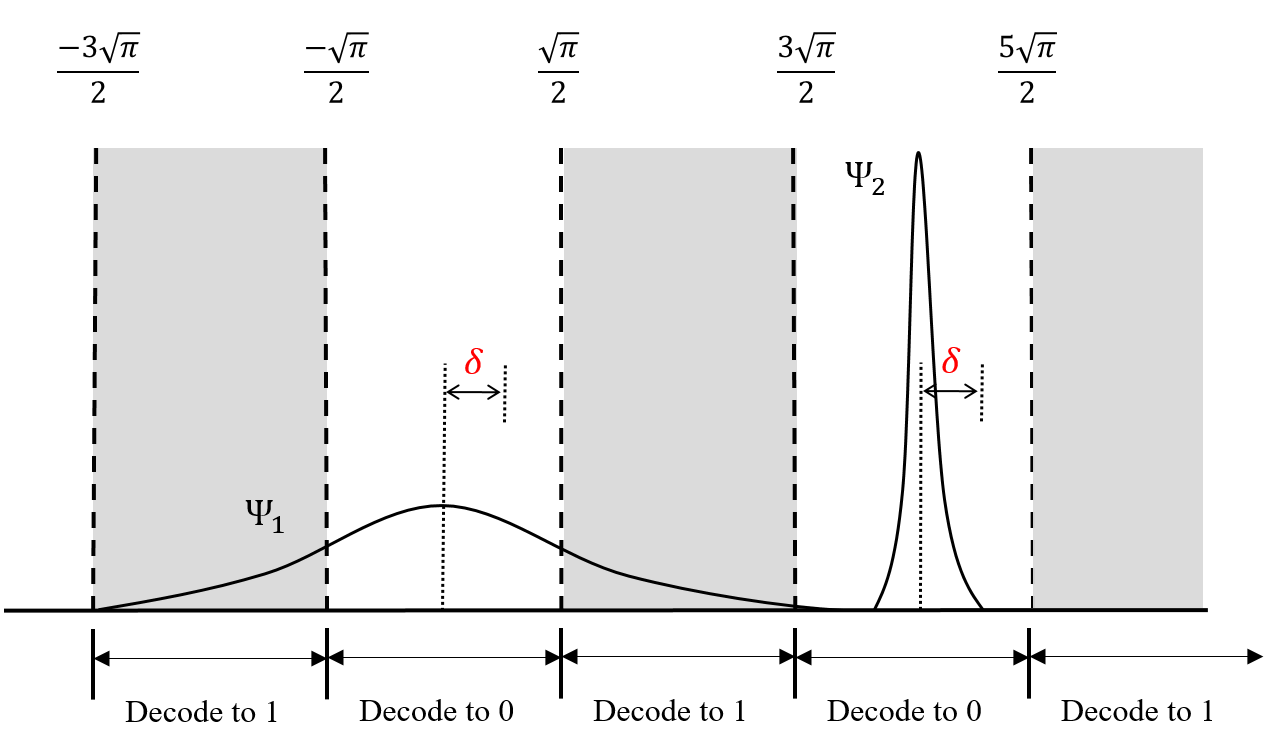}
\caption{Peaks centered at even integer multiples of $\sqrt{\pi}$ in $q$ space. The peak on the left contains large tails that extend into the region where a shift error is decoded to the logical $\ket{\overline{1}}$ state. The peak on the right is much narrower. Consequently for some interval $\delta$, the peak on the right will correct shift errors of size $\frac{\sqrt{\pi}}{2} - \delta$ with higher probability than the peak on the left.}
\label{fig:IllustrationPeaks}
\end{figure}

Naturally because of the finite width of the peaks of approximate GKP states, it will not be possible to correct a shift error in $p$ or $q$ of magnitude at most $\frac{\sqrt{\pi}}{2}$ with certainty. For example, suppose we have an approximate $\ket{\overline{0}}$ GKP state with a peak at $q=0$ subject to a shift error $e^{-ivp}$ with $|v| \le \frac{\sqrt{\pi}}{2}$. The finite width of the Gaussian peaks will have a non-zero overlap in the region $\frac{\sqrt{\pi}}{2} < q < \frac{3 \sqrt{\pi}}{2}$ and $\frac{-3 \sqrt{\pi}}{2} < q < \frac{-\sqrt{\pi}}{2}$. Thus with non-zero probability the state can be decoded to $\ket{\overline{1}}$ instead of $\ket{\overline{0}}$ (see \cref{fig:IllustrationPeaks} for an illustration). 

In general, if an approximate GKP state is afflicted by a correctable shift error, we would like the probability of decoding to the incorrect logical state to be as small as possible. A smaller overlap of the approximate GKP state in regions in $q$ and $p$ space that lead to decoding the state to the wrong logical state will lead to a higher probability of correcting a correctable shift error by a perfect GKP state. 
 
 \subsubsection{Preparation of approximate GKP states using Phase Estimation} 
\begin{figure}
    \centering
    \includegraphics[width=0.43\textwidth]{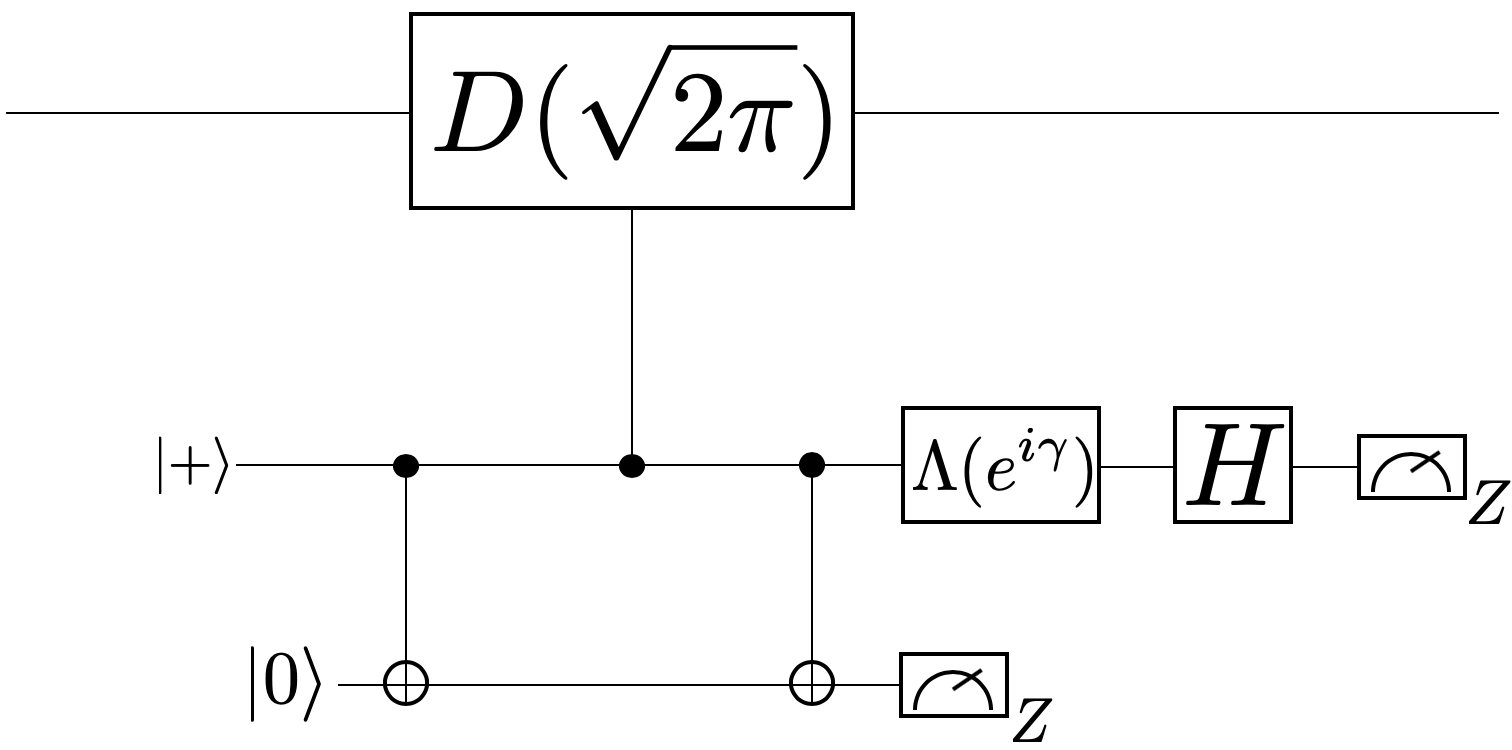}   
    \caption{Phase estimation circuit with the flag qubit. The protocol is aborted if the flag qubit measurement is non-trivial.}
    \label{fig:ControlDispAmpDampFig}
\end{figure}

We observe that the GKP states are the eigenstates of the $S_p$ operator, thus we can use phase estimation to gradually project a squeezed vacuum state to an approximate GKP state. The phase estimation circuit for preparing an approximate $\ket{\tilde{0}}$ GKP state is given in \cref{fig:ControlDispAmpDampFig}. The first horizontal line represents the cavity mode that we want to prepare the GKP states. The second line is a transmon ancilla initialized to $\ket{+}$. The third line is a transmon flag qubit initialized to$\ket{0}$. The $H$ gate is the Hadamard gate. $\Lambda(e^{i \gamma}) = \text{diag}(1,e^{i\gamma})$ is the gate with a control parameter $\gamma$ in each round of the phase estimation to . After applying several rounds of the circuit in \cref{fig:ControlDispAmpDampFig}, the input squeezed vacuum state is projected onto an approximate eigenstate of $S_p$ with some random eigenvalue $e^{i\theta}$. Additionally, an estimated value for the phase $\theta$ is obtained. After computing the phase, the state can be shifted back to an approximate $+1$ eigenstate of $S_p$.

In our protocol, we use a flag qubit to detect any damping event during in the controlled-displacement gate, if a non-trivial measurement is obtained, we abort the protocol and start over. Using our simulation results, we also find a subset of output states that are robust to measurement errors in the transmon ancilla and only accept states in that subset. We proved that our protocol is fault-tolerant according to the definition we gave. In practice, our protocol produces ``good" approximate GKP states with high probability and we expect to see experimental efforts to implement our protocol. 
 
\subsection{Discussion}

The GKP qubit architecture is a promising candidate for both near-term and fault-tolerant quantum computing implementations. With intrinsic error correcting capabilities, the GKP qubit breaks the abstraction layer between error correction and the physical implementation of qubits. In \cite{shiNJP2019}, we discussed the faul-tolerant preparation of GKP qubits and realistic experimental difficulties. We believe that qubit encodings like the GKP encoding will be useful for  reliable quantum computing.
\section{Conclusion and Future Directions}

In this review, we proposed that greater quantum efficiency can be achieved by breaking abstraction layers in the quantum computing stack. We examined some previous work in this direction that are closing the gap between current quantum technology and real-world quantum computing applications. We would also like to briefly discuss some promising future directions along this line.

\subsection{Noise-Tailoring Compilation} We can further explore the idea of breaking the ISA abstraction. Near-term quantum devices have errors from elementary operations like 1- and
2-qubit gates, but also emergent error modes like cross-talk. Emergent error modes are hard to characterize and to mitigate. Recently, it has been shown that randomized compiling could transform complicated noise channels including cross-talk, SPAM errors and readout errors into simple stochastic Pauli errors \cite{Wallman_2016}, which could potentially enable subsequent noise-adaptive compilation optimizations. We believe if compilation schemes that combine noise tailoring and noise adaptation could be designed, they will outperform existing compilation methods.

    \subsection{Algorithm-Level Error Correction} Near-term quantum algorithms such as Vairiational Quantum Eigensolver (VQE) and Quantum Approximate Optimization Algorithm (QAOA) are tailored for NISQ hardware, breaking the circuit/ISA abstraction. We could take a step further and look at high level algorithms equipped with customized error correction/mitigation schemes. Prominent examples of this idea are the Generalized Superfast Encoding (GSE) \cite{setia2018superfast} and the Majorana Loop Stabilizer Code (MLSC) \cite{jiang2018majorana} for quantum chemistry. In GSE and MLSC, the overhead of mapping Fermionic operators onto qubit operators stays constant with the  qubit number (as opposed to linear scaling in the usual Jordan-Wigner encoding or logarithmic in Bravyi-Kitaev encoding). On the other hand, qubit operators in these mappings are logical operators of a distance 3 stabilizer error correction code so that we can correct all weight 1 qubit errors in the algorithm with stabilizer measurements. These work are the first attempts to  algorithm-level error correction and we are expecting to see more efforts of this kind to improve the robustness of near-term algorithms.

\subsection{Dissipation-Assisted Error Mitigation}  We generally think of dissipation as competing with quantum coherence. However, with careful design of the quantum system, dissipation can be engineered and used for improving the stability of the underlying qubit state. Previous work on autonomous qubit stabilization \cite{yao17} and error correction \cite{Kapit_2016} suggest that properly engineered dissipation could largely extend qubit coherence time. Exploring the design space of such systems and their associated  error correction/mitigation schemes might provide  alternative paths to an efficient and scalable quantum computing stack.

\section*{Acknowledgements}
This work is funded in part by EPiQC, an NSF Expedition in Computing, under grants CCF-1730449/1832377/1730082; in part by STAQ, under grant NSF Phy-1818914; and in part by DOE grants DE-SC0020289
and DE-SC0020331.

\bibliographystyle{plain}
\bibliography{references}
\end{document}